\documentclass[aps,prl,preprint,superscriptaddress]{revtex4}

\usepackage{graphicx}
\usepackage{dcolumn}
\usepackage{bm}
\usepackage{hyperref}
\usepackage{verbatim}
\usepackage{color}
\usepackage{csquotes}

\begin{document}
\newcommand{\lao}{LaAlO$_3$} 
\newcommand{\sto}{SrTiO$_3$} 
\newcommand{\lasto}{(La,Al)$_{1-x}$(Sr,Ti)$_x$O$_3$} 
\newcommand{\nthreed}{$n_\mathrm{3D}$}
\newcommand{\nthreedth}{$n_\mathrm{3D,th}$} 
\newcommand{\ntwod}{$n_\mathrm{2D}$} 
\newcommand{\ttwog}{$t_\mathrm{2g}$}
\newcommand{\dxy}{$d_{xy}$}
\newcommand{\dyz}{$d_{yz}$}
\newcommand{\dxz}{$d_{xz}$}
\newcommand{\dxzyz}{$d_{xz/yz}$}
\newcommand{\SI}{\textit{Supporting Information}}

\title{Probing Quantum Confinement and Electronic Structure at Polar Oxide Interfaces}

\author{D.~Li}
 \email[]{denverli@stanford.edu}
 \thanks{Present address: Department of Applied Physics, Stanford University, Stanford, California 94305, USA}
 \affiliation{Department of Quantum Matter Physics, University of Geneva, 24 quai Ernest-Ansermet, CH-1211 Geneva 4, Switzerland}
 \author{S.~Lemal}
 \affiliation{Theoretical Materials Physics, Q-MAT, CESAM, Universit\'e de Li\`ege, B-4000 Li\`ege, Belgium}
 \author{S.~Gariglio}
 \affiliation{Department of Quantum Matter Physics, University of Geneva, 24 quai Ernest-Ansermet, CH-1211 Geneva 4, Switzerland}
  \author{Z.~P.~Wu}
 \affiliation{Department of Quantum Matter Physics, University of Geneva, 24 quai Ernest-Ansermet, CH-1211 Geneva 4, Switzerland}
 \affiliation{State Key Laboratory of Information Photonics and Optical Communications and School of Science, Beijing University of Posts and Telecommunications, Beijing 100876, China}
   \author{A.~F\^ete}
 \affiliation{Department of Quantum Matter Physics, University of Geneva, 24 quai Ernest-Ansermet, CH-1211 Geneva 4, Switzerland}
   \author{M.~Boselli}
 \affiliation{Department of Quantum Matter Physics, University of Geneva, 24 quai Ernest-Ansermet, CH-1211 Geneva 4, Switzerland}
 \author{Ph.~Ghosez}
 \affiliation{Theoretical Materials Physics, Q-MAT, CESAM, Universit\'e de Li\`ege, B-4000 Li\`ege, Belgium}
   \author{J.-M.~Triscone}
 \affiliation{Department of Quantum Matter Physics, University of Geneva, 24 quai Ernest-Ansermet, CH-1211 Geneva 4, Switzerland}

\date{\today}
\maketitle

{\bf
Polar discontinuities occurring at interfaces between two different materials constitute both a challenge and an opportunity in the study and application of a variety of devices. In order to cure the large electric field occurring in such structures, a reconfiguration of the charge landscape sets in at the interface via chemical modifications, adsorbates or charge transfer. In the latter case, one may expect a local electronic doping of one material: one sparkling example is the two-dimensional electron liquid (2DEL) appearing in \sto\ once covered by a polar \lao\ layer. Here we show that tuning the formal polarisation of a \lasto\ (LASTO:$x$) overlayer through chemical composition modifies the quantum confinement of the 2DEL in \sto\ and its electronic band structure. The analysis of the behaviour in magnetic field of superconducting field-effect devices reveals, in agreement with \textit{ab initio} calculations and self-consistent Poisson-Schr\"odinger modelling, that quantum confinement and energy splitting between electronic bands of different symmetries strongly depend on interface charge densities. These results not only strongly support the polar discontinuity mechanisms with a full charge transfer to explain the origin of the 2DEL at the celebrated \lao/\sto\ interface, but also demonstrate an effective tool for tailoring the electronic structure at oxide interfaces.


}





Functionalities offered by the interfaces between different materials have been the fundamental basis of modern electronic devices and information technology \cite{SchenkelNatMat2005,SzeBook2006}. Transition metal oxides (TMOs), largely owing to their correlated $d$ electrons and entanglement of various degrees of freedom, provide an ideal platform for establishing a variety of novel electronic properties \cite{TokuraScience2000,CheongNatMat2007}. In recent years, a burst of research activities on heterostructures of transition metal oxides has been driven by the quest for emergent phenomena at interfaces that are absent in the bulk parent compounds \cite{HwangNatMatRev2012}. Indeed, at the interface, different phenomena are at play, including symmetry breaking, electrostatic coupling, atomic rearrangement, etc.~\cite{ZubkoAnnRev2011,ChakhalianRevModPhys2014}. One phenomenon particularly relevant for transition metal ions with multiple valence states is charge transfer. This effect has been suggested to be the origin of several recently discovered novel electronic states: its driving force has been attributed to differences in cationic electron affinity for instance in LaNiO$_3$/LaMnO$_3$ \cite{GibertNatMat2012,LeePRB2013} or CaMnO$_3$/CaRuO$_3$ \cite{TakahashiAPL2001,HePRL2012}, or to charge delocalisation in LaTiO$_3$/SrTiO$_3$ \cite{OhtomoNature2002,OkamotoNature2004} or LaMnO$_3$/SrMnO$_3$ \cite{MayNatMat2009,SmadiciPRL2007} heterostructures. 

For the \lao\ (LAO)/\sto\ (STO) interface, the polar discontinuity has often been regarded as the origin of the charge transfer \cite{OhtomoNature2004,NakagawaNatMat2006,ThielScience2006,LeiNPJ-QM2017} and the formation of a superconducting two dimensional electron liquid (2DEL) \cite{ReyrenScience2007,CavigliaNature2008}. The formal polarisation of LAO due to the (LaO)$^+$ and (AlO$_2$)$^-$ atomic planes is disrupted along the [001] direction by the non-polar STO, since (SrO)$^0$ and (TiO$_2$)$^0$ planes are formally charge neutral.
Density-functional theory (DFT) calculations \cite{PopovicPRL2008,PentchevaJPCM2010} for a defect-free interface show that the build-up of the electric potential inside LAO leads to a Zener breakdown \cite{NakagawaNatMat2006} and a progressive charge transfer of 0.5 electrons per surface unit cell (0.5~e$^-$/u.c.) from the LAO surface O-$2p$ states to the interfacial Ti-$3d$ states, thus forming a 2DEL. This mechanism successfully explains the observed threshold LAO thickness, $t_\mathrm{c}$, of 4 unit cells (u.c.) for the onset of conductivity  \cite{ThielScience2006}, but faces questions raised by several experiments, reporting, for instance, the lack of O-$2p$ hole-pockets \cite{BernerPRL2013} or a weak initial electric field in LAO \cite{BernerPRB2013,SegalPRB2009,SlootenPRB2013}. Relying on polar discontinuity and the formation of oxygen vacancies at the LAO surface, recent theoretical work \cite{BristowePRB2011,BristoweJPCM2014review,YuZungerNatComm2014,ZhouPRB2015} predicts a similar critical thickness $t_\mathrm{c}$, while solving some of the discrepancies mentioned above.

Reinle-Schmitt \textit{et al.} \cite{ReinleNatComm2012} showed that polar discontinuity and critical thickness are indeed correlated: tuning the polar discontinuity by alloying LAO with STO [\lasto, denoted as LASTO:$x$], $t_\mathrm{c}$ increases, with an inverse proportionality to the fraction of LAO in the alloy and therefore with the formal polarisation, in perfect agreement with DFT predictions.
The self-confinement of the transferred charge leads to an electronic reconstruction of the Ti $t_\mathrm{2g}$ bands, increasing the energy of \dxzyz-symmetry states by $\sim$ 50~meV with respect to the \dxy\ states \cite{SalluzzoPRL2009,DelugasPRL2011}. This band structure differs from the electronic structure of bulk STO \cite{vanderMarelPRB2011}, and is at the origin of the unique electronic properties of the LAO/STO interface. However, the carrier density (\ntwod) measured using Hall effect at the LASTO:0.5/STO interface was found to be comparable to the one of the standard LAO/STO interface (of the order of $\sim$ 10$^{13}$~cm$^{-2}$) \cite{ReinleNatComm2012}, a value significantly lower than the theoretical prediction based on the polar discontinuity scenario (\ntwod = 1.7$\times$10$^{14}$~cm$^{-2}$ or 0.25~e$^-$/u.c. for the LASTO:0.5/STO interface and 3.3$\times$10$^{14}$~cm$^{-2}$ or 0.5~e$^-$/u.c. for the LAO/STO interface). This discrepancy between the predicted and measured carrier densities has been one important open issue questioning the origin of the 2DEL and has been related to different mechanisms, such as charge localization \cite{PopovicPRL2008,TakizawaPRB2011,CancellieriStrocovBook2018} due to interface disorder, surface defects and selective quantum confinement in a single layer, or to anti-site defects \cite{YuZungerNatComm2014}, or to phase separation \cite{AriandoNatComm2011PhaseSeparation}.

Here we probe the quantum confinement and its consequences on the electronic band structure at the superconducting LASTO:0.5/STO interface and compare the results with standard LAO/STO interfaces. Using sophisticated superconductivity measurements and DFT-complemented Poisson-Schr\"odinger modelling, we provide strong evidences that, although transport measures a fraction of the carriers predicted by the polar discontinuity scenario, the full amount of charge is indeed present at the interfaces. The field-effect tuning of the superconducting state suggests that the energy splitting between bands of different symmetries (\dxy\ vs. \dxzyz) is reduced with the widening of the confining potential, in agreement with the models. Our results strongly support that charges with nominal carrier densities of 0.5~e$^-$/u.c. and 0.25~e$^-$/u.c. for the LAO/STO and LASTO:0.5/STO interfaces respectively contribute to the electronic confinement and only these total carrier densities can explain the differences in the extension of the 2DEL and in their electronic configurations.

\begin{figure}[p]
	\centering
	\includegraphics[width= \textwidth]{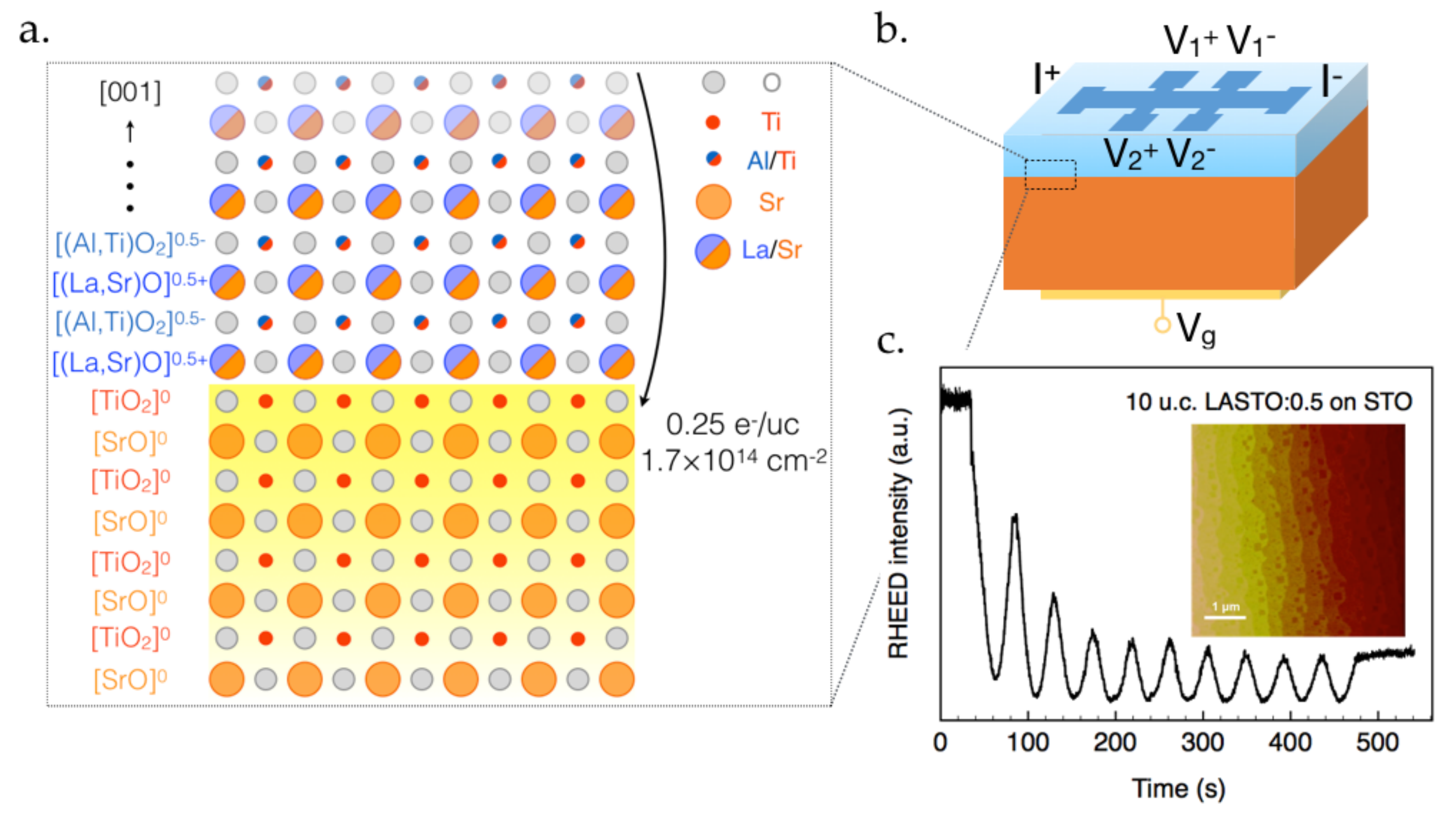}
	\caption{\textbf{Schematics of the LASTO:0.5/STO structure and sample preparation.} \textbf{a.} A detailed illustration of the interfacial structure with atomic arrangements and charges per atomic plane. Electrons with surface density of 0.25~e$^-$/u.c.\ are expected to be transferred. \textbf{b.} A sketch of the LASTO:0.5/STO field-effect device (5$\times$5~mm$^2$) with the back-gate electrode used to apply the gate bias ($V_\mathrm{g}$) facing the 500~$\mu{m}$-wide Hall-bar. \textbf{c.} Oscillations of \textit{in-situ} RHEED intensity during the growth of a 10~u.c.-thick LASTO:0.5 layer and (inset) its surface topography acquired by atomic force microscopy (AFM), revealing the step-and-terrace morphology.}
	\label{Fig1}
\end{figure}

Figure~\ref{Fig1}a shows a schematic of the atomic structure of the LASTO:0.5/STO interface with the La/Sr and Al/Ti cations spread evenly on the \textit{A}O and \textit{B}O$_2$ perovskite planes according to energy calculations performed by DFT \cite{ReinleNatComm2012}. Consequently, along the $[0 0 1]$ direction, the layer has alternating planes with formal electronic charges of $+0.5$ and $-0.5$ per unit cell (u.c.). The polar discontinuity is therefore half the one of standard LAO/STO interfaces. As a result, the density of the transferred charges to the interface is estimated to be 0.25~e$^-$/u.c.\ ($\sim$ 1.7$\times$10$^{14}$~cm$^{-2}$). LASTO:0.5 films were grown by pulsed-laser deposition (PLD) using growth conditions described in \textit{Experimental Section}. The oscillations of the reflection high-energy electron diffraction (RHEED) intensity provide a measure of the thickness of the layer during the growth, as shown in Figure~\ref{Fig1}c for a 10~u.c. film. Topographic scan using atomic force microscopy (AFM) shows a sharp step-and-terrace structure on the film surface with single-unit-cell step height (Figure~\ref{Fig1}c inset). Both RHEED and AFM data indicate a high growth quality of the films. We note that, in order to probe the intermixing strength and the sharpness of the LASTO:0.5/STO interface, detailed scanning transmission electron microscopy (STEM) study is required. Hall bars (see Figure~\ref{Fig1}b) were defined using a pre-patterning technique for transport measurements \cite{CancellieriEPL2010}. A gate electrode was deposited at the backside of the STO substrate (see Figure~\ref{Fig1}b) in order to tune the properties of the electron liquid. Transport measurements were performed in a dilution cryostat equipped with a rotating sample holder allowing the orientation of the magnetic field to be varied from perpendicular to parallel to the interface plane.

\begin{figure}[ht!]
	\centering
	\includegraphics[width= \textwidth]{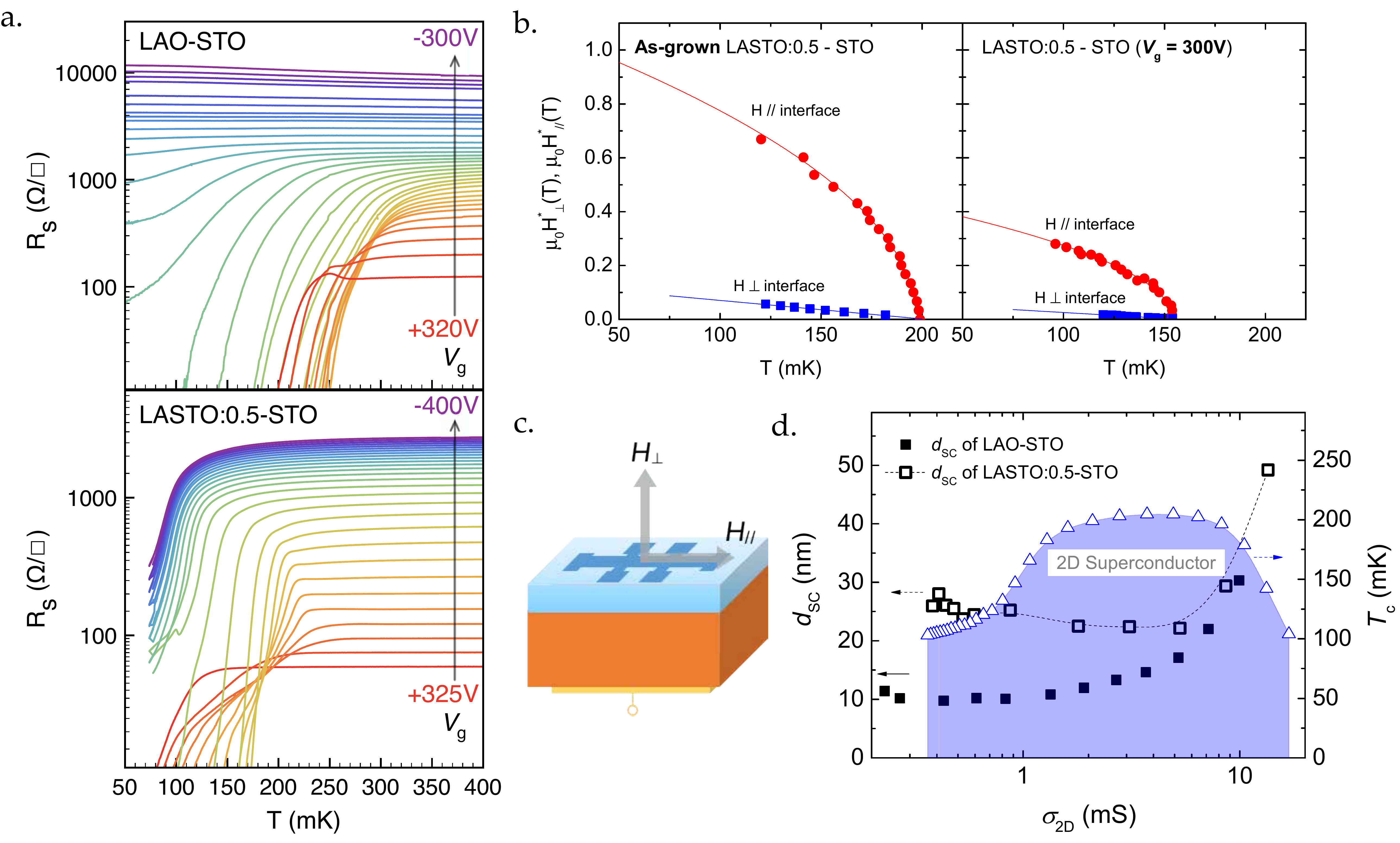}
	\caption{\textbf{Superconducting properties and phase diagram of the LASTO:0.5/STO interface.} \textbf{a.} Field-effect modulation of sheet resistance ($R_\mathrm{s}$) as a function of $T$ on a semi-logarithmic scale of the LASTO:0.5/STO interface (\textbf{bottom} panel), in comparison with the LAO/STO interface (\textbf{top} panel, replotted from \cite{CavigliaNature2008}). For LASTO:0.5/STO, the gate voltage varies from $-400$~V to $+325$~V in steps of $25$~V. \textbf{b.} Perpendicular (dots) and parallel (squares) critical fields ($H^*_{\perp}$, $H^*_{\parallel}$) as a function of temperature ($T$) for the as-grown state (\textbf{left} panel) and for a state with $V_\mathrm{g} = 300$~V (\textbf{right} panel) for LASTO:0.5/STO. The solid line for $H^*_{\parallel}$ is a fit to the data using the Ginzburg-Landau formula for a 2D film (see text), while the line for $H^*_{\perp}$ is a guide to the eye. \textbf{c.} Configurations for the measurements of superconducting critical fields applied perpendicular and parallel to the interface plane. \textbf{d.} Phase diagram of the LASTO:0.5/STO interface, showing in blue the superconducting state, plotted as a function of the normal-state sheet conductance. The superconducting layer thickness $d_\mathrm{SC}$ (left axis, black open square) is compared to that of the LAO/STO interface \cite{GariglioAPLMater2016} (left axis, black solid square). The black dotted line is a guide to the eye.}
	\label{Fig2}
\end{figure}

A superconducting ground state is observed when the sample is cooled below a critical temperature, $T_\mathrm{c}$, of $\sim$ 200~mK, generally lower than that of the LAO/STO system (see discussions) \cite{CavigliaNature2008}. In Figure~\ref{Fig2}b we show the critical magnetic field $H^*(T)$ for perpendicular and parallel orientation of a LASTO:0.5/STO interface in its virgin state. The critical temperature for each magnetic field was defined as the temperature at which the sheet resistance $R_\mathrm{s}$ reaches half the normal state value (estimated at 500~mK). We calculate the in-plane coherence length $\xi_\parallel (T)$ from the Ginzburg-Landau $H_{c2}$ formula using the perpendicular field (Equation \ref{eq1}) and we extrapolate for $T = 0$ a value of $\sim$ 55~nm, similar to the one observed for standard interfaces \cite{ReyrenAPL2009}.
\begin{equation}
	\xi_\mathrm{GL}(0)=\sqrt{\frac{\Phi_0}{2\pi\mu_{0}H_{c2,\perp(0)}}}
	\label{eq1}
\end{equation}
The temperature dependence of the parallel critical field $H^*_\parallel(T)$ follows the 2D behaviour of a superconducting thin film $H^*_\parallel(T)$ $\propto$ $[1-T/T_\mathrm{c}(H=0)]^{1/2}$ \cite{TinkhamBook1996}. We then use Equation \ref{eq2} for a 2D superconductor to calculate its thickness from the value of $H^*_\parallel(T)$:
\begin{equation}
	d_\mathrm{SC}=\frac{\sqrt{3}\Phi_0}{\pi\xi_\parallel(T)\mu_{0}H^*_\parallel(T)}
	\label{eq2}
\end{equation}
We obtain a value $d_\mathrm{SC} \sim$ 24~nm, independent of the temperature as expected, a value that is larger by a factor of 2 than the one of LAO/STO interfaces \cite{CopiePRL2009,DubrokaPRL2010,ReyrenAPL2009}. We note that the thickness obtained from the analysis of the anisotropic superconducting properties is a characteristic thickness which, in the case of pure LAO/STO, was found to agree well with the vertical spread of the electron system determined using other methods \cite{CopiePRL2009,DubrokaPRL2010}.

Field effect was then used to tune the doping level and the superconducting properties of the interface. Figure~\ref{Fig2}a shows the modulation of the superconducting transitions when the gate voltage is varied from $-400$~V to $+325$~V, in comparison with the data of the LAO/STO interface \cite{CavigliaNature2008}. We observe that $T_\mathrm{c}$ and the normal state resistance (see \SI\ for the $R_\mathrm{s}$ vs.\ $V_\mathrm{g}$ plot) are effectively tuned; we note however that for the largest negative voltages, i.e. in the strongest depletion regime, the system remains metallic and superconducting and we do not attain the insulating state reported for standard LAO/STO interfaces (top panel in Figure~\ref{Fig2}a). The presence of superconductivity and relatively high conductance at largest negative voltages can be related to the occupation of \dxzyz-symmetry states, even at lowest doping levels, at the LAOSTO:0.5/STO interface and will be discussed below. The evolution of $T_\mathrm{c}$ versus $\sigma_\mathrm{2D}$ (sheet conductance in the normal state), illustrated in Figure~\ref{Fig2}d, shows a dome-like behaviour, reaching a maximum of 206 mK. For several gate voltages, the parallel and perpendicular critical magnetic fields were measured: data for $V_\mathrm{g} = 300$~V is shown in the right panel of Figure~\ref{Fig2}b. These measurements allow us to estimate the superconducting layer thickness across the phase diagram. This information is shown in Figure~\ref{Fig2}d, where the superconducting thickness for LASTO:0.5/STO samples is displayed in comparison with LAO/STO interfaces. We see that the thickness for the alloy samples is larger than that of standard interfaces across the phase diagram: $d_\mathrm{SC}$ of LASTO:0.5/STO interfaces falls in the range of 20-30~nm for a wide interval of conductances, while it remains 10-15~nm for the LAO/STO interface. Assuming that this superconducting thickness mirrors an effective width of the confining potential, we attribute this enhancement to the difference in the transferred charge due to a modified polar discontinuity.

To support this idea, we model the quantum confinement at these oxide interfaces using two complementary approaches, first-principles calculations and Poisson-Schr\"odinger (P-S) approach \cite{JanottiPRB2012}.

Previous DFT studies on LAO/STO interfaces \cite{DelugasPRL2011,LeePRB2008,YamamotoAPL2014,JanottiPRB2012,ReshakJAP2016,CancellieriPRB2014,GuJPCC2012} have investigated the fully compensated interface (3.3 $\times$ 10$^{14}$~cm$^{-2}$) through symmetric superlattices (with two $n$-type interfaces). Lower carrier densities have been reached by artificially removing some charges at the interface and compensating by a positive background \cite{DelugasPRL2011, CancellieriPRB2014}. In our study, the interfacial charge is directly modified by explicitly tuning the polar discontinuity, i.e. by switching LAO to LASTO:0.5 solid solution. In the following, we present the two doping cases, 0.5~e$^-$/u.c.\ and 0.25~e$^-$/u.c.\ for (LAO)$_{2}$/(STO)$_{30}$ and (LASTO:0.5)$_{2}$/(STO)$_{30}$ superlattices (each with 330 atoms, see \SI) . For the purpose of describing the charge profile at the interface, we set the extension of the STO block as wide as possible for numerical calculations (much larger than previous studies \cite{DelugasPRL2011,CancellieriPRB2014}). The calculations also provide physical parameters used later in the P-S modelling.

\begin{figure}[ht!]
	\centering
	\includegraphics[width=0.7 \textwidth]{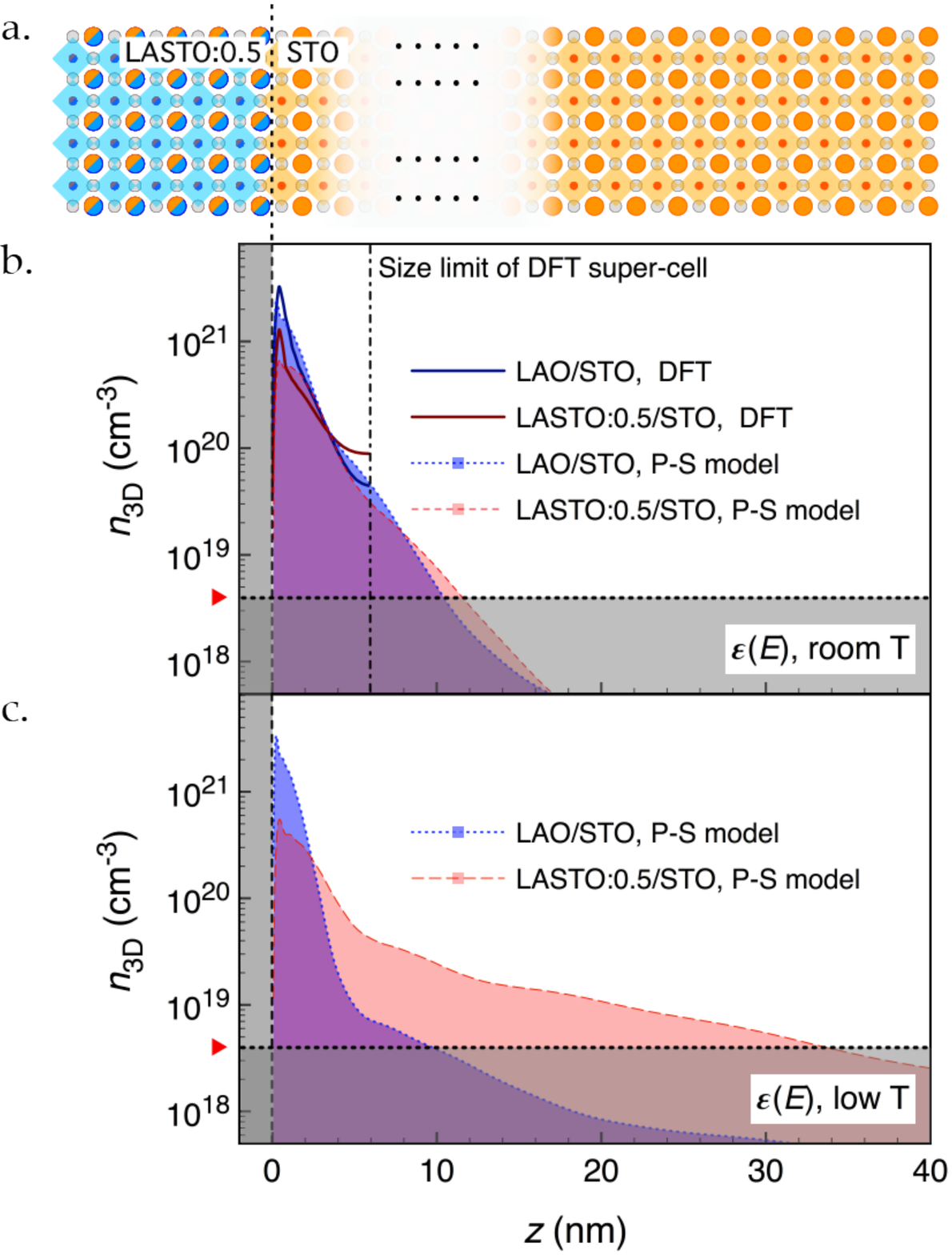}
	\caption{\textbf{Charge distributions in STO calculated for the two interfaces.} The panels display the charge density (\nthreed) in logarithmic scale as a function of distance ($z$) from the interface for \ntwod\ = 0.5~e$^-$/u.c.~(LAO/STO) and 0.25~e$^-$/u.c.~(LASTO:0.5/STO). \textbf{a.} Schematic of atomic structures of the LASTO:0.5/STO interface. \textbf{b.} Results from DFT calculations (solid lines) and P-S model (dotted/dash lines) for a room-temperature field-dependent STO dielectric constant $\epsilon(E)$ (see \cite{StengelPRL2011} and \SI). \textbf{c.} Charge profile estimated with the P-S model (dotted/dash line) using the low-temperature $\epsilon(E)$ (\SI). Black dash line indicates the interface. Dot-dashed line corresponds to the size limit of the largest supercell used in DFT calculations. Dotted lines and the red triangles indicate the 3D density threshold ($n_\mathrm{3D,th} \approx$ 4$\times$10$^{18}$~cm$^{-3}$) for the occurrence of superconductivity in STO.}
	\label{Fig3}
\end{figure}

Figure~\ref{Fig3} illustrates a sketch of the LASTO:0.5/STO system (Figure~\ref{Fig3}a) and compares the main results from the calculations. The dashed line indicates the position of the interface. The DFT computed charge profile $n_\mathrm{3D} (z)$ for the two interfaces is displayed in Figure~\ref{Fig3}b. The vertical dot-dashed line denotes the extension of the largest STO supercell. Performing calculations for different sizes of the STO supercell, we have noticed that the charge profile extends further into the substrate each time we increase the supercell width, but this occurs on a logarithmic charge density scale (see \SI). Looking at the plot, we see that most of the carriers are located in the first unit cells in both cases; moving away from the interface, a crossing between the two profiles occurs (at $z \approx$ 3.5~nm): for lower carrier density, the charge is more spread into the supercell while for higher density it is more localized close to the interface. These calculations rely on the self-consistent DFT dielectric constant ($\epsilon_\mathrm{STO}$ = 250 in zero field) that reflects the behaviour of STO at room temperature.

To access the low temperature case which is experimentally measured, we need to overcome some limitations of the DFT approach: (i) the STO size limit in the numerical calculations and (ii) the low temperature diverging value of the dielectric constant $\epsilon_\mathrm{STO}$ that is not captured. To bypass these limitations, a self-consistent P-S model using DFT data is employed \cite{SternPRB1972}. Effective mass $m^*$ values are obtained from DFT calculations. We notice that the effect of the renormalization of $m^*$ due to, for instance, polaronic effects \cite{DubrokaPRL2010,CancellieriNatComm2016}, has also been examined (See \SI): while this renormalization indeed slightly affects the calculated charge distribution, it does however not alter the main findings, discussions and conclusions below. The field dependence of the dielectric constant $\epsilon_\mathrm{STO} (E)$ is adjusted according to the measurements \cite{PeelaersAPL2015} (details can be found in \textit{Experimental Section} and \SI).

The resulting profile for the room-temperature configuration ($\epsilon_\mathrm{STO}$ = 250) is first compared with the DFT data in Figure~\ref{Fig3}b: the two approaches provide similar charge density profiles (especially in the regime close to the interface), therefore validating the P-S model with respect to the DFT calculations; it further highlights that, although large and at the computation limit, the size of the DFT supercells is still too small to correctly describe the tail of the carrier density profile, which is better accessible in the P-S simulation. From the latter we determine that for both surface carrier densities, the spatial extension at the 3D density threshold for the occurrence of superconductivity ($n_\mathrm{3D,th} \approx$ 4$\times$10$^{18}$~cm$^{-3}$, labelled in Figure~\ref{Fig3}b) \cite{KooncePR1967STO} is $\sim$10~nm. We note that the thickness of the 2DEL estimated by different techniques differs a lot -- from a few u.c.~to $\sim$10~nm \cite{BasleticNatMat2008,CopiePRL2009,SingPRL2009,CancellieriPRL2013}. This may be due to the high density in the first few atomic layers of STO that is decaying by a factor of 10 in 5~nm. The tail of this distribution may fall below the sensitivity of some of the techniques used to probe the 2DEL thickness.

At low temperature, $\epsilon_\mathrm{STO}$ increases due to the quantum paraelectric behaviour of STO: in this regime, the field dependence is described by Equation~\ref{eq3} (see \textit{Experimental Section} and Figure~S3 for details). Using this dependence, the charge density profiles \nthreed($z$) are calculated using the P-S model for the two interfaces and are illustrated in Figure~\ref{Fig3}c. We see that the effect of the large $\epsilon_\mathrm{STO}$ is to open the confining potential for the LASTO:0.5/STO interface with more charges spreading into STO. At the \nthreedth\ (labelled in Figure~\ref{Fig3}c), the extension of the 2DEL is $\sim$ 10~nm for the LAO/STO interface and 31~nm for the LASTO:0.5/STO interface, values that agree with the estimation of the superconducting thickness extracted from the critical field measurements discussed above. This suggests that the change in superconducting thickness and the confinement scale at the LASTO:0.5/STO interface is due to a reduced total charge density originating from a modified polar discontinuity (see discussion below). The modelling also clearly shows that the dielectric properties of STO and the total 2D carrier density define the confinement of the charge, in agreement with a recent tight-binding study \cite{RaslanPRB2017}. We note that, at low temperature, the P-S modelling gives a lower integrated charge density for \nthreed($z$) above \nthreedth (blue area above the dotted line in Figure~\ref{Fig3}c as compared to that in Figure~\ref{Fig3}b). This is also due to the remarkable temperature dependence of $\epsilon_\mathrm{STO} (T)$ and the large electric field dependence of $\epsilon_\mathrm{STO} (E)$ at low temperature: a portion of the charges extends further into STO, with density below \nthreedth.

\begin{figure}[ht!]
	\centering
	\includegraphics[width=0.95 \textwidth]{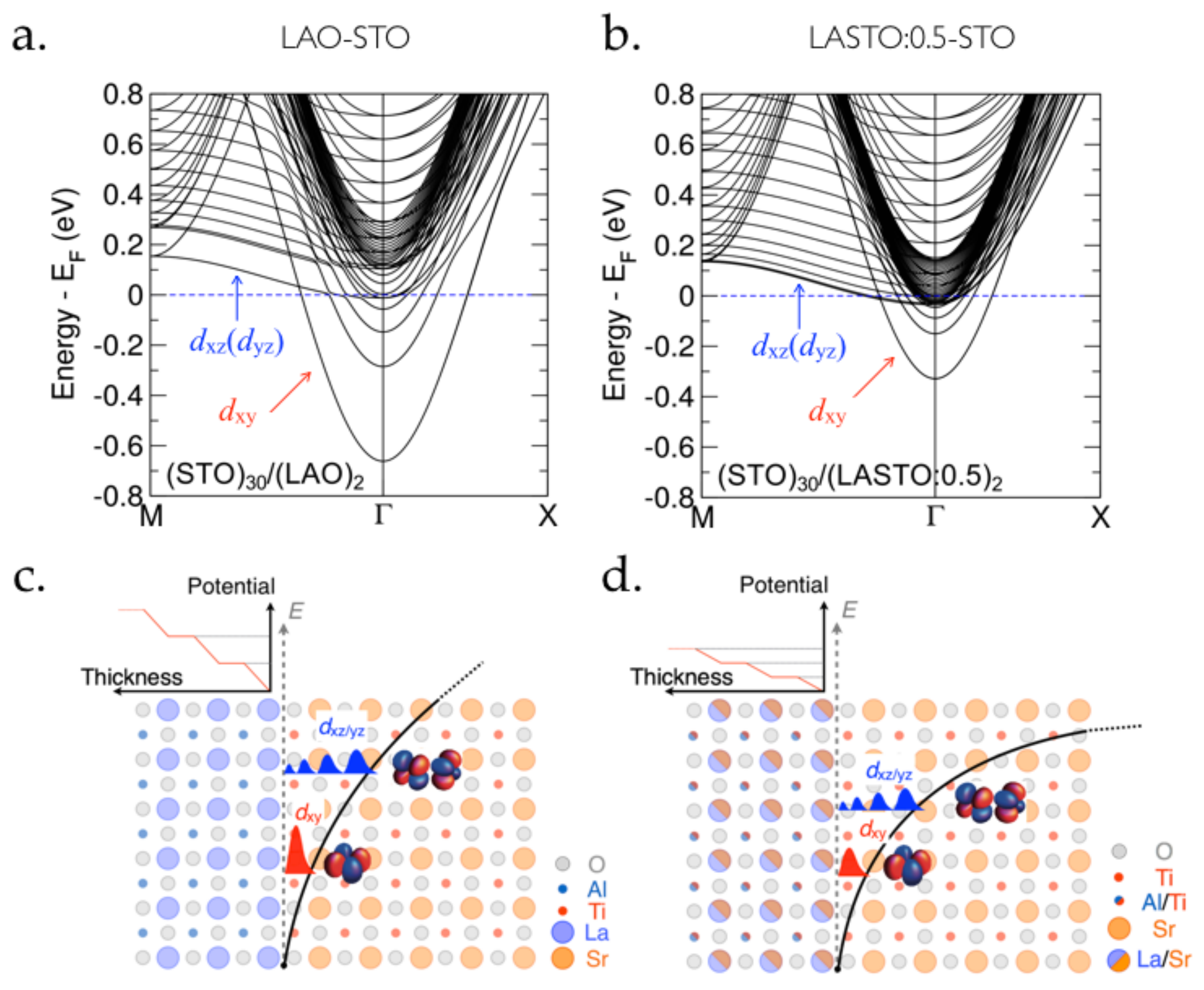}
	\caption{\textbf{Electronic structure of the two interfaces.} Band structures of the LAO/STO (\textbf{a.}) and the LASTO:0.5/STO (\textbf{b.}) interfaces, calculated by DFT. The \ttwog\ bands are labelled according to their symmetry (\dxy, \dxzyz). Schematics of the atomic arrangements, \textit{built-in} electric potential (red lines), quantum confinement potential (black lines) and \dxy\ - \dxzyz\ band splittings of the LAO/STO (\textbf{c.}) and the LASTO:0.5/STO (\textbf{d.}) interfaces. For a larger extension of the 2DEL, the band splitting is reduced.}
	\label{Fig4}
\end{figure}

The effect of the surface carrier density and the quantum confinement on the electronic band structures calculated from DFT for the two superlattices (STO)$_{30}$/(LAO)$_{2}$ and (STO)$_{30}$/(LASTO:0.5)$_{2}$ is displayed in Figure~\ref{Fig4}, panels a and b. The Fermi level lies at $0~eV$ and the symmetry of the electronic states (\dxy\ and \dxzyz) is indicated by an arrow. We see changes in the splittings between sub-bands as well as between bands with different symmetry: for a larger extension of the 2DEL (LASTO:0.5/STO, panel b), the sub-band spacing is reduced and the \dxz/\dyz\ bands are occupied at lower densities. The calculated bottom of the first \dxz/\dyz\ band relative to Fermi level depends on the size of the superlattices, which is discussed in \SI\ (Figure S8).


The agreement between the experimental estimation of the superconducting layer thickness and the theoretical calculations provides strong support to the polar discontinuity as the doping mechanism of the interfaces. Despite the similar mobile carrier densities extracted from Hall measurements for the two interfaces \cite{ReinleNatComm2012}, calculations and superconducting critical fields measurements show that the increase in the characteristic thickness of the 2DEL for the LASTO:0.5/STO interface is a direct result of the reduced \textit{total} transferred charges (from 0.5 to 0.25~e$^-$/u.c.) upon modification of the polar discontinuity. Although a large fraction of these charges does not contribute to electric transport, it defines the strength of the electric field in STO and consequently its dielectric constant at the interface, that finally sets the extension of the 2DEL. In fact, considering only a charge density extracted from Hall measurements ($\sim$ 0.05~e$^-$/u.c.) at the interfaces in P-S calculations would result in an ineffective electric field and de-confinement of the electrons with the spatial extension over $\sim$ 60~nm at density of \nthreedth. This demonstrates that all charges with nominal charge density predicted by the polar discontinuity mechanism are transferred to the interface. Our results show significant differences from the work by Ueno et al. \cite{UenoPRB2014}, where, experimentally, no dependence of the superconducting thickness on the electrostatically-tuned carrier density was observed.

The consequence of the larger charge profile at the LASTO:0.5/STO interface is a band structure with larger contribution of the \dxzyz\ bands to the density of states, as shown in Figure~\ref{Fig4}b. We argue that this electronic configuration is the reason for the modified phase diagram of the superconducting state of the LASTO:0.5/STO interface presented in Figure~\ref{Fig2}. The overall lower $T_\mathrm{c}$ observed in the system, as compared with that of the LAO/STO interface, can be due to a shallow confining potential, which gives a reduced effective \nthreed\ as the extension increases \cite{GariglioAPLMater2016,ValentinisPRB2017}. Furthermore, we observe that field effect is not efficient enough to deplete the 2DEL in order to suppress superconductivity nor to reach a highly resistive state. This is clearly different from the LAO/STO system. The presence of superconductivity and the low resistance state in the depletion regime suggest that the \dxzyz-symmetry states are occupied even at low dopings due to reduced energy splittings between bands, highlighting the relevance of these orbitals on superconductivity and high electron mobility at the interfaces.

Our study provides a direct comparison between \textit{ab initio} calculations performed on a massive supercell and the self-consistent P-S modelling, revealing that the latter method is also valid to describe quantum confinement of oxide systems where $3d$ electronic states are more spatially localised than $p$ bands in semiconductor. The \textit{ab initio} calculations of the effective masses combined with the experimental determination of the dielectric properties allow the P-S modelling to describe the field and charge configuration on scales that today cannot be achieved by DFT methods.

To conclude, we have demonstrated that by doping LAO films with STO to form a 50~\% alloy compound we are able to successfully change the polar discontinuity at this (super-)conducting interface. This modulation leads to a significant change in the interfacial confining potential that we have estimated from the measurements of the characteristic superconducting thickness. The evolution of the confinement with the change in polarization is captured by advanced large-scale DFT calculations and self-consistent Poisson-Schr\"odinger modelling, only when assuming that a full transfer of charges takes place at the LAO/STO interface with the density predicted by the polar catastrophe model using first-principle calculations. The resulting band structure for the LASTO:0.5/STO interface reveals a larger contribution of the \dxzyz\ bands to the density of states and explains the persistence of the superconducting state in the depletion regime. This study shows that the control of the polar discontinuity at oxide interfaces by chemical composition is an effective tool for engineering novel electronic states in these compounds.

\section{EXPERIMENTAL SECTION}
\subsection{Sample preparation}
Prior to film deposition, Hall-bar patterns with the crystalline TiO$_2$-terminated surface of (001) single-crystal STO substrates were defined using photolithography, followed by a subsequent amorphous STO deposition used as hard mask and lift-off process. Films were then grown by pulsed-laser deposition using standard growth conditions \cite{FeteAPL2015}: a KrF laser ($248$~nm) with a pulse energy of $40$~mJ ($\approx 0.8$~J\,cm$^{-2}$), repetition rate of $1$~Hz; growth temperature of $800~^{\circ}$C, O$_2$ pressure of $1\times10^{-4}$~mbar; samples were cooled after growth to $540~^{\circ}$C in $200$~mbar O$_2$ and maintained at this temperature and pressure for one hour before being cooled down to room temperature in the same atmosphere. The growth rate was approximately $55$~laser pulses per monolayer. The deposition was fully monitored by RHEED and specular spot intensity oscillations indicate a layer-by-layer growth. RBS analysis reveals a film stoichiometry in agreement, within the experimental uncertainties (1.5~\%), with the nominal concentration of $x = 0.5$ for a $40$-nm-thick film.

\subsection{Field-effect device and superconductivity measurements}
Aluminium wires were ultra-sonically bonded to the sample. Gold pad was deposited by sputtering on the back side of the sample as back-gate electrode. A dc bias was applied across the STO substrate between back gate and the conducting interface. Superconductivity measurements were performed in a $^3$He/$^4$He dilution refrigerator (Leiden Cryogenics) with a base temperature of 50~mK and a superconducting magnet allowing field of up to 15~T to be reached. Samples were attached to a rotator head for anisotropic magnetic field measurements. Precise parallel and perpendicular orientations with respect to the magnetic field were determined using both longitudinal resistance $R_\mathrm{xx}$ and Hall resistance $R_\mathrm{xy}$. In the parallel direction, an \textit{off-axis} angle is estimated to be smaller than 0.02$^{\circ}$ from the $R_\mathrm{xy}$ signal. Current was applied to the conducting channel from a Keithley 6220 high precision current source. Voltages were recorded using Keithley 2182 nanovolt meters.

\subsection{DFT calculations}
{First-principles calculations have been performed with the CRYSTAL code \citep{refcrystal14}, which implements the Kohn-Sham ansatz to density functional theory~\cite{kohnsham} using a Linear Combination of Atomic Orbital (LCAO) approach and local gaussian basis sets. Electronic exchange-correlation effects were described with the B1-WC hybrid functional~\cite{dbilcb1wc}, including $16\%$ of Hartree-Fock exact exchange. Sampling of the Brillouin zone with a Monkhorst-Pack~\cite{monkhorstpack1972} $6 \times 6 \times 1$ k-point mesh ensures a proper convergence of the total energy at the self-consistent field level, with a threshold criterium of $\rm 10^{-8}~Ha$. The electronic properties are computed using a refined $12 \times 12 \times 2$ k-point mesh. A gaussian smearing of the Fermi surface has been set to $\rm 0.001~Ha$. The basis sets used for different atoms are detailled in Ref.~\cite{tibasisset} for Ti, Ref.~\cite{osrbasisset} for O, Ref.~\cite{albasisset} for Al and Ref.~\cite{labasisset} for La, and correspond to the ones used in Ref.~\cite{ReinleNatComm2012}.}
{The optimisation of the atomic positions is performed with convergence criteria of $1.5 \times 10^{-4}\rm~Ha/Bohr$ in the root-mean square values of the energy gradients, and $1.2 \times 10^{-3}\rm~Bohr$ in the root-mean square values of the atomic displacements. The evaluation of the Coulomb and exchange series is determined by five parameters, fixed to their default~\cite{refcrystal14manual} values: 7, 7, 7, 7 and 14.}
{Calculations are performed on off-stoichiometric $\rm (STO)_{30}/(LAO)_{2}$ and $\rm (STO)_{30}/(LASTO$:$0.5)_{2}$ superlattices, with two \emph{n}-type interfaces, with an additional $\rm TiO_{2}$ plane in STO, and an additional $\rm LaO$ ($\rm LaSrO$) plane in LAO ($\rm LASTO$).}
The effective masses associated to the different Ti \ttwog\ band are calculated. Each \ttwog\ band has two light effective masses associated to the light carriers and one heavy effective mass associated to the heavy carriers (for Ti \dxy, $m_{xx}^{*} = m_{yy}^{*} = m_{L}^{*}$ and $m_{zz}^{*} = m_{H}^{*}$). With both superlattices, we have $m_{L}^{*}$ = 0.4~$m_\mathrm{e}$ and $m_{H}^{*}$ = 5.9~$m_\mathrm{e}$ (very close to the calculated values in cubic STO: $m_{L}^{*}$ = 0.4~$m_\mathrm{e}$ and $m_{H}^{*}$ = 6.1~$m_\mathrm{e}$).

\subsection{P-S modelling}
Starting from the band structure at the LAO/STO interface suggested by DFT calculations \cite{DelugasPRL2011}, we model the confinement, energy levels, and charge distribution at both interfaces by self-consistently solving the Poisson and Schr\"odinger equations \cite{SternPRB1972}. For the boundary conditions of the potential profile $V(z)$, we consider that the LAO or the LASTO:0.5 layer fixes the displacement field $D$ at the interface to be $D$ = \ntwod$e$ with \ntwod\ = 0.5 or 0.25~e$^-$/u.c., respectively, while $D$ = 0 at the bottom side of STO. The static dielectric constant of STO calculated by DFT ($\epsilon_\mathrm{STO}^\mathrm{DFT} (E = 0)$ = 250) was used as the room-temperature static $\epsilon_\mathrm{STO} (E = 0)$. The low-temperature static $\epsilon_\mathrm{STO} (E = 0)$ was obtained from a fit to our experimental data ($\epsilon_\mathrm{STO} (E = 0)$ = 25462). For both temperatures, the high-field dependent $\epsilon_\mathrm{STO} (E)$ is estimated from Stengel \cite{StengelPRL2011}.  $\epsilon_\mathrm{STO} (E)$ takes the general form:
\begin{equation}
	\epsilon_\mathrm{STO}(E) = 1 + \frac{B}{[1 + (E/E_\mathrm{0})^2]^{1/3}}
	\label{eq3}
\end{equation}
with $B$ = 250, $E_\mathrm{0}$ = 10$^7$~V m$^{-1}$ for room temperature, and $B$ = 25462, $E_\mathrm{0}$ = 82213~V m$^{-1}$ for low temperature, respectively (see \SI).


For details of the calculation, the algorithm used for convergence of the self-consistency calculations, please refer to \cite{FeteThesis2014}. The field dependence of $\epsilon_\mathrm{STO} (E)$ can be found in \SI.

\section{Acknowledgments}

Danfeng Li and S\'ebastien Lemal contributed equally to this work. We thank C. Cancellieri for providing the target and M. Gabay for useful discussions. This work was supported by the Swiss National Science Foundation -- division II, and has received funding from the European Research Council under the European Union Seventh Framework Programme (FP7/2007-2013)/ERC Grant Agreement No.\ 319286 (Q-MAC). The present research also benefits from computational resources made available on the Tier-1 supercomputer of the F\'ed\'eration Wallonie-Bruxelles, infrastructure funded by the Walloon Region under the grant agreement $\rm n^{\circ}$ 1117545. Computational resources have been provided by the Consortium des \'Equipements de Calcul Intensif (C\'ECI), funded by the Fonds de la Recherche Scientifique de Belgique (F.R.S.-FNRS) under Grant No.\ 2.5020.11. Ph.~G. and S.~L. acknowledge financial support from F.R.S.-FNRS PDR project HiT-4FiT and from ARC project AIMED.



\bibliographystyle{Classes/naturemag}
\bibliography{References.bib}

%

\end{document}


\newcommand{\lao}{LaAlO$_3$} 
\newcommand{\sto}{SrTiO$_3$} 
\newcommand{\lasto}{(La,Al)$_{1-x}$(Sr,Ti)$_x$O$_3$} 
\newcommand{\nthreed}{$n_\mathrm{3D}$} 
\newcommand{\ntwod}{$n_\mathrm{2D}$} 
\newcommand{\ttwog}{$t_\mathrm{2g}$}
\newcommand{\dxy}{$d_{xy}$}
\newcommand{\dyz}{$d_{yz}$}
\newcommand{\dxz}{$d_{xz}$}
\newcommand{\dxzyz}{$d_{xz/yz}$}
\newcommand{\SI}{\textit{Supplementary Information}}

\setcounter{equation}{0}
\setcounter{figure}{0}
\setcounter{table}{0}
\setcounter{page}{1}
\makeatletter
\renewcommand{\theequation}{ES\arabic{equation}}
\renewcommand{\thetable}{S\arabic{table}}
\renewcommand{\thefigure}{S\arabic{figure}}

\title{{SUPPORTING INFORMATION for} \\ Probing Quantum Confinement and Electronic Structure at Polar Oxide Interfaces}

\author{D.~Li}
 \email[]{denverli@stanford.edu}
 \thanks{Present address: Department of Applied Physics, Stanford University, Stanford, California 94305, USA}
 \affiliation{Department of Quantum Matter Physics, University of Geneva, 24 quai Ernest-Ansermet, CH-1211 Geneva 4, Switzerland}
 \author{S.~Lemal}
  \thanks{Contributed equally to this work}
 \affiliation{Theoretical Materials Physics, Q-MAT, CESAM, Universit\'e de Li\`ege, B-4000 Li\`ege, Belgium}
 \author{S.~Gariglio}
 \affiliation{Department of Quantum Matter Physics, University of Geneva, 24 quai Ernest-Ansermet, CH-1211 Geneva 4, Switzerland}
  \author{Z.~P.~Wu}
 \affiliation{Department of Quantum Matter Physics, University of Geneva, 24 quai Ernest-Ansermet, CH-1211 Geneva 4, Switzerland}
 \affiliation{State Key Laboratory of Information Photonics and Optical Communications and School of Science, Beijing University of Posts and Telecommunications, Beijing 100876, China}
   \author{A.~F\^ete}
 \affiliation{Department of Quantum Matter Physics, University of Geneva, 24 quai Ernest-Ansermet, CH-1211 Geneva 4, Switzerland}
   \author{M.~Boselli}
 \affiliation{Department of Quantum Matter Physics, University of Geneva, 24 quai Ernest-Ansermet, CH-1211 Geneva 4, Switzerland}
 \author{Ph.~Ghosez}
 \affiliation{Theoretical Materials Physics, Q-MAT, CESAM, Universit\'e de Li\`ege, B-4000 Li\`ege, Belgium}
   \author{J.-M.~Triscone}
 \affiliation{Department of Quantum Matter Physics, University of Geneva, 24 quai Ernest-Ansermet, CH-1211 Geneva 4, Switzerland}

\maketitle

\subsection{Field Effect}

\begin{figure}[h!]
	\centering
	\includegraphics[width=0.75 \textwidth]{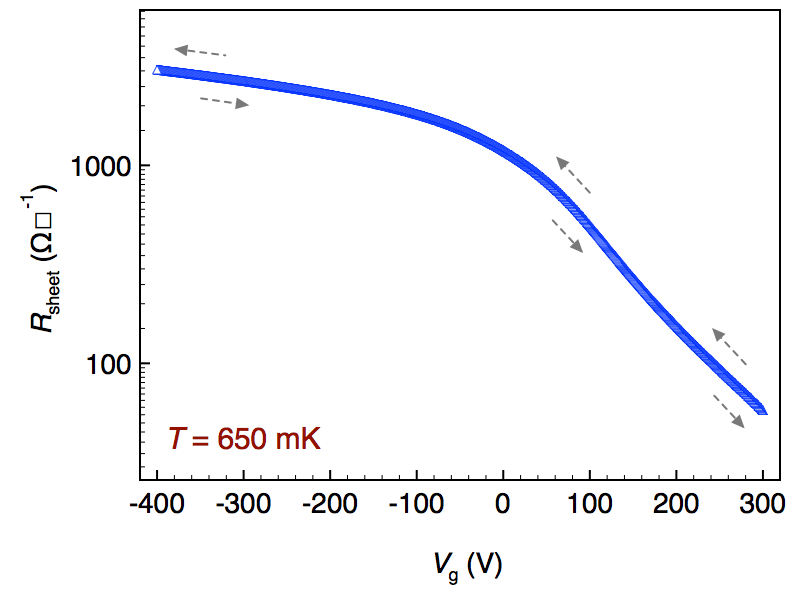}
	\caption{Back-gate modulation of sheet resistance at the LASTO:0.5/STO interface. The curve is obtained after a \enquote{training} process by sweeping the gate voltage, $V_\mathrm{g}$, from $0$~V to $+300$~V at 650~mK. The dashed arrows indicate a reversible behaviour in $V_\mathrm{g}$ range of $+300$~V to $-400$~V.}\label{fig:RsvsVg}\end{figure}

A \enquote{training} procedure was applied prior to gating experiments to avoid hysteresis behaviour in sheet resistance $R_\mathrm{s}$ as a function of back-gate voltage $V_\mathrm{g}$ \cite{CavigliaNature2008,LiuAPLMater2015}: $V_\mathrm{g}$ was firstly swept from $0$~V to the maximum positive voltage, $+300$~V. After the \enquote{training} step, as shown in Fig.~\ref{fig:RsvsVg}, a reversible $R_\mathrm{s}$ vs. $V_\mathrm{g}$ behaviour can be observed. The curve was taken at 650~mK. We can see that for negative $V_\mathrm{g}$ (depletion regime), the change in $R_\mathrm{s}$ is not dramatic, as compared to the LAO-STO interface.

\subsection{P-S Model}

\subsubsection{Technical details}

The general procedure of the self-consistent Poisson-Schr\"odinger model is briefly described below. The band mass and envelope function approximations are used. Parameters concerning carriers from the $n$-th sub-band generated by the quantum confinement of the bulk band $\alpha$ are calculated. The indexed parameters used in the calculations are listed in Table~\ref{parametersforPSmodel}.

\begin{table}[htb]
\centering
\caption{Definitions of the parameters used in the Poisson-Schr\"odinger calculation for different sub-band $n$ generated by the quantum confinement of the bulk band $\alpha$}
\vspace{10px}
\begin{tabular}[c]{|c|c|}
\hline\hline
&Definition\\
\hline
$m^{*,\alpha}_i$&Band mass for the bulk band $\alpha$ along $i$ direction\\
\hline
$\xi^{\alpha}_n$&Amplitude envelope function\\
\hline
$\epsilon^{\alpha}_n$&Solution of Eq.~\ref{equa:Schrodinger} along $\hat{z}$\\
\hline
$E^{\alpha}_0$&Energy bottom of the bulk band $\alpha$\\
\hline
$E^{\alpha}_n$&Energy levels of different sub-bands\\
\hline\hline
\end{tabular}
\label{parametersforPSmodel}
\end{table}

A triangular potential well $V(z)$ that corresponds to a constant electric field in STO is taken as a starting point. The energy levels $E^{\alpha}_n$ for each band $\alpha$ in the well can be calculated using the two Eqs.~\ref{equa:Schrodinger} and \ref{equa:energylevels}:

\begin{equation}
-\frac{\hbar^2}{2m^{*,\alpha}_z}\frac{\partial^2\xi^{\alpha}_n(z)}{\partial z^2} + eV(z)\xi^{\alpha}_n(z) + E^{\alpha}_0\xi^{\alpha}_n(z) = \epsilon^{\alpha}_n\xi^{\alpha}_n(z)
\label{equa:Schrodinger}
\end{equation}

\begin{equation}
E^{\alpha}_n(\mathbf{k_\parallel}) = \frac{\hbar^2k^2_x}{2m^{*,\alpha}_x} + \frac{\hbar^2k^2_y}{2m^{*,\alpha}_y} + \epsilon^{\alpha}_n
\label{equa:energylevels}
\end{equation}where Eq.~\ref{equa:Schrodinger} is the Schr\"odinger equation. The solution of this equation $\epsilon^{\alpha}_n$ can be used to calculate the energy levels. Furthermore, the Fermi energy $E_\mathrm{F}$ can be determined by Eq.~\ref{equa:Ef} setting the total sheet carrier density equal to $n_\mathrm{2D}$. This input value is 0.5 and 0.25~e$^-$/u.c. for LAO/STO and LASTO:0.5/STO interfaces, respectively.

\begin{equation}
n_\mathrm{2D} = \sum_{n,\alpha} \Theta (E_\mathrm{F} - \epsilon^{\alpha}_n)\frac{\sqrt{m^{*,\alpha}_xm^{*,\alpha}_y}}{\pi\hbar^2} (E_\mathrm{F} - \epsilon^{\alpha}_n)
\label{equa:Ef}
\end{equation}with $\Theta$ the heaviside function. With this information the charge density $\rho_\mathrm{3D}$ is calculated using Eq.~\ref{equa:3Dprofile}:

\begin{equation}
\rho_\mathrm{3D}(z) = e\sum_{n,\alpha} \Theta (E_\mathrm{F} - \epsilon^{\alpha}_n)\frac{\sqrt{m^{*,\alpha}_xm^{*,\alpha}_y}}{\pi\hbar^2} (E_\mathrm{F} - \epsilon^{\alpha}_n) {\xi^{\alpha}_n(z)}^2
\label{equa:3Dprofile}
\end{equation} and is used in the Poisson equation (\ref{equa:Poisson}) to compute a new potential well that fits the charge distribution profile.

\begin{equation}
-\frac{\partial}{\partial z}(\epsilon_0\epsilon_r(E)\frac{\partial V(z)}{\partial z}) = \rho_\mathrm{3D}(z)
\label{equa:Poisson}
\end{equation}This potential well $V(z)$ will then be re-injected into Eq.~\ref{equa:Schrodinger}. Multiple iterations are performed until a defined convergence criterion is satisfied (see Ref.~\cite{FeteThesis2014} for details). As a result, the potential well is directly related to the sheet carrier density $n_\mathrm{2D}$. We also note that, in Eq.~\ref{equa:Poisson} enters the electric field dependence of the STO dielectric constant, $\epsilon_\mathrm{STO} (E)$, which we discuss below.

\subsubsection{Dielectric constant of STO}

\begin{figure}[ht!]
	\centering
	\includegraphics[width=0.75 \textwidth]{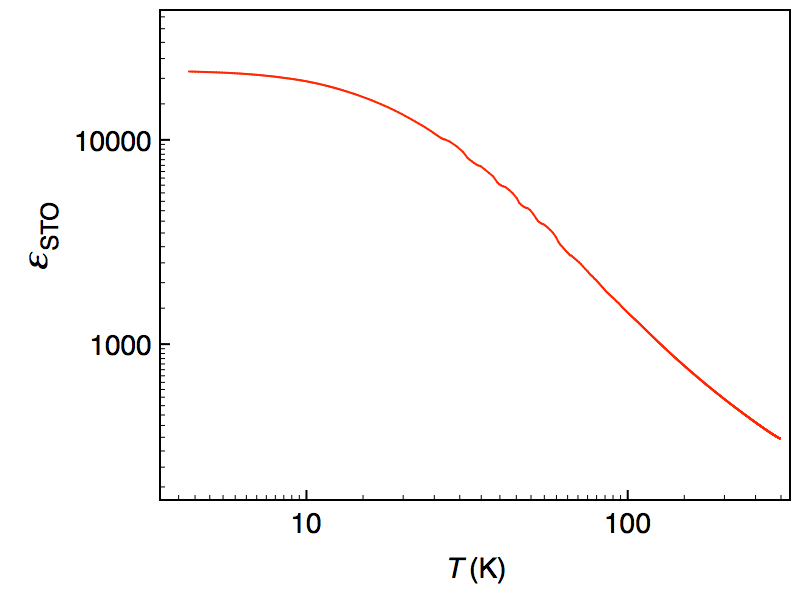}
	\caption{Temperature evolution of the dielectric constant of STO, $\epsilon_\mathrm{STO} (T)$, measured on STO single crystal using differential capacitance technique.}\label{fig:Tdependepsilon}\end{figure}
The temperature dependence of the dielectric constant of STO single crystal (0.5~mm thick),  $\epsilon_\mathrm{STO}$, was measured and is shown in Fig.~\ref{fig:Tdependepsilon}. Same as reported \cite{SakudoPRL1971,MullerPRB1979}, STO shows behaviour of an incipient ferroelectric as $\epsilon_\mathrm{STO}$ increases dramatically upon cooling, especially at low temperatures and plateaus at $\sim$ 10$^{4}$ when approaching to 0~K. This remarkable enhancement of $\epsilon_\mathrm{STO}$ affects the confining potential at conducting oxide interfaces.

\begin{figure}[ht!]
	\centering
	\includegraphics[width=0.75 \textwidth]{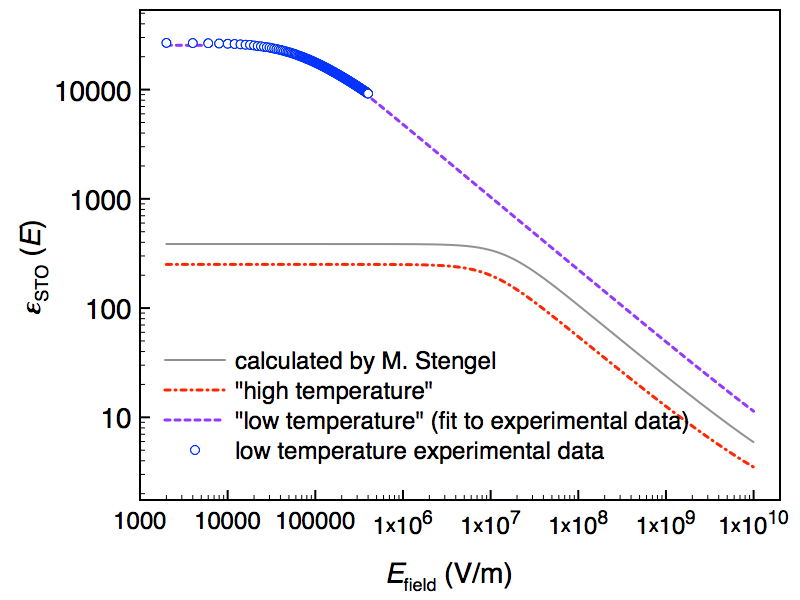}
	\caption{Field dependence of the dielectric constant of STO, $\epsilon_\mathrm{STO} (E)$, used in P-S modelling. \textit{Grey line}: calculation by M. Stengel \cite{StengelPRL2011}; \textit{red dot-dashed line}: calculated data following \cite{StengelPRL2011} with a fixed low field $\epsilon_\mathrm{STO}$, which is obtained by DFT calculation (with equation used in Methods); \textit{blue open circle}: experimental data measured at 4~K; \textit{purple dash line}: fitting to low temperature experimental data using equation in Methods.}\label{fig:fielddependsilon}\end{figure}

The electric field dependence of $\epsilon_\mathrm{STO}$ is shown in Fig.~\ref{fig:fielddependsilon}. As discussed in the main text, the static $\epsilon_\mathrm{STO}$ ($\sim$ 250) calculated by DFT corresponds to a room temperature low field value. The field dependence, $\epsilon_\mathrm{STO} (E)$, in this case, follows Eq.~3 shown in the main text and the calculation by Stengel \cite{StengelPRL2011}. $\epsilon_\mathrm{STO} (E)$ at low temperature has been experimentally measured (blue open circles in Fig.~\ref{fig:fielddependsilon}). A fit to the data gives $\epsilon_\mathrm{STO} (E)$ for the whole field range at low temperature, which also takes a similar form (Eq.~3 in the main text).

\subsubsection{Confining potential}

\begin{figure}[ht!]
	\centering
	\includegraphics[width=0.75 \textwidth]{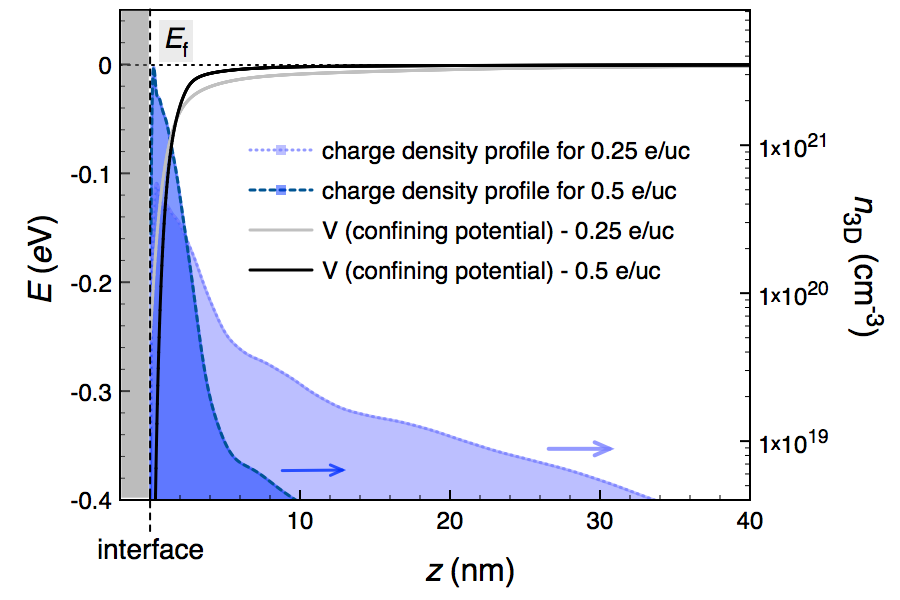}
	\caption{Confining potential profiles for the low temperature case calculated by P-S model. \textit{Right axis}: charge density profile for low temperature $\epsilon_\mathrm{STO} (E)$, same as in Fig.~3b; \textit{left axis}: corresponding confining potential renormalised by Fermi energy.}\label{fig:confiningpotentiallowT}\end{figure}

A large discrepancy in characteristic thickness and charge distribution for LASTO:0.5/STO and LAO/STO interfaces using low temperature $\epsilon_\mathrm{STO} (E)$ has been illustrated in Fig.~3b in the main text. We plot here the corresponding confining potential profiles for both interfaces (0.25~e$^-$/u.c. and 0.5~e$^-$/u.c.). From the results we can see that the effect of charge de-confinement mostly become dramatic at the top of the confining well. This indicates that it is a small portion (compared to the total amount) of the transferred charges, which populates bands close to Fermi energy, that contributes to the large difference in 2DEL thickness.

\subsubsection{Effective mass}

\begin{figure}[ht!]
	\centering
	\includegraphics[width=0.7 \textwidth]{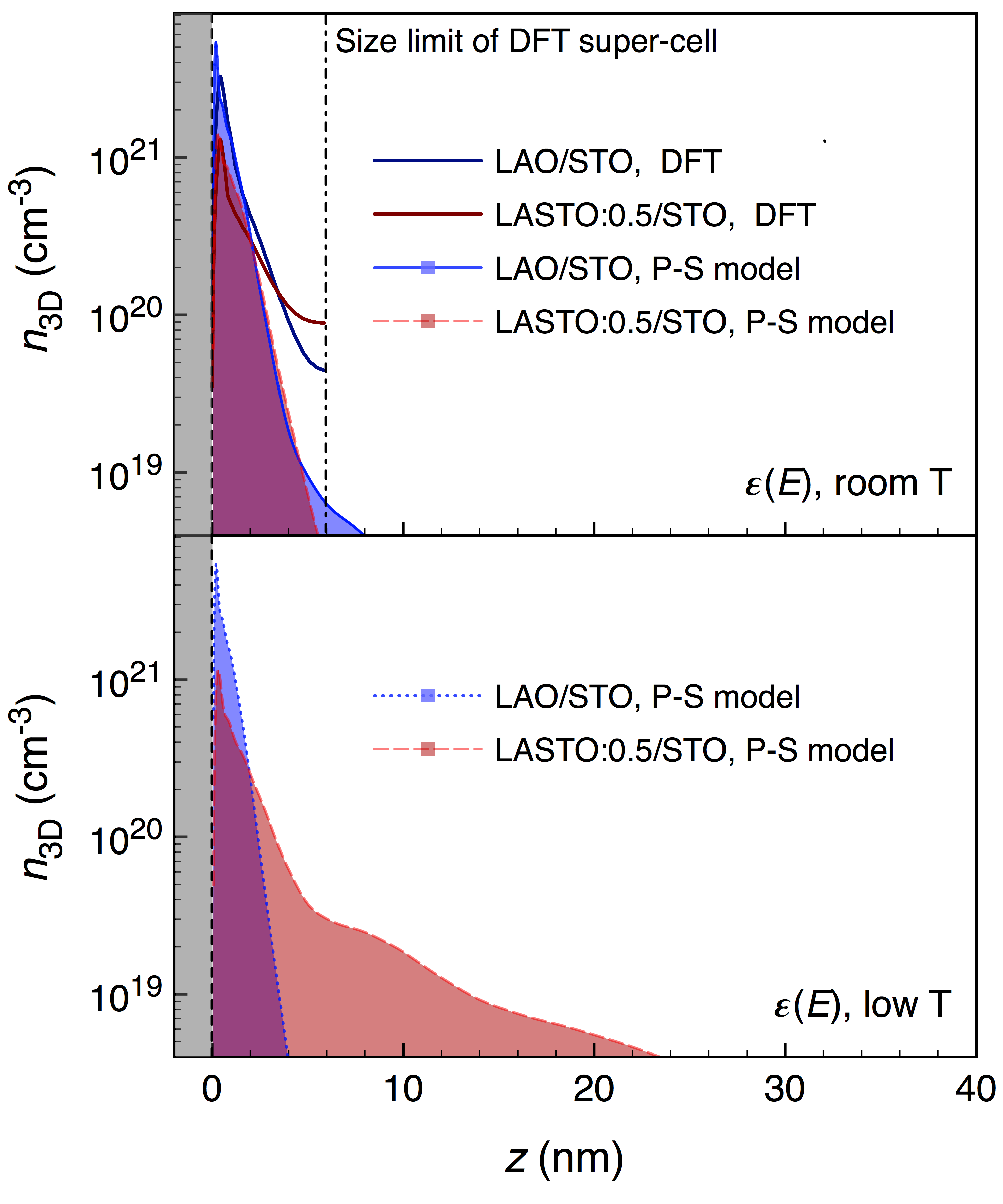}
	\caption{Charge distributions in STO calculated for the two interfaces using renormalized $m^*$ (by a factor of 3.5). The panels display the charge density (\nthreed) in logarithmic scale as a function of distance ($z$) from the interface for \ntwod = 0.5 e$^-$/u.c. (LAO/STO) and 0.25 e$^-$/u.c. (LASTO:0.5/STO). \textit{Top panel}: Results from DFT calculations (solid lines) and P-S model (dotted/dash lines) for a room-temperature field-dependent STO dielectric constant $\epsilon_\mathrm{STO} (E)$. \textit{Bottom panel}: Charge profile estimated with the P-S model (dotted/dash lines) using the low-temperature $\epsilon_\mathrm{STO} (E)$. Black dash line indicates the interface. Dot-dashed line corresponds to the size limit of the largest supercell used in DFT calculations.}\label{fig:n3DwithmPolaronic}\end{figure}

To examine the effect of the effective mass $m^*$ on the P-S modeling results, we performed the P-S calculation using a renormalized $m^*$ (by a factor of 3.5, as compared to $m^*$ obtained from DFT), which can originate from polaronic effect \cite{CancellieriNatComm2016,DubrokaPRL2010}. The resulted charge distributions $n_\mathrm{3D}(z)$ using both "room-temperature" and "low-temperature" $\epsilon_\mathrm{STO} (E)$ are plotted in Fig.~\ref{fig:n3DwithmPolaronic}, as compared to Figure 3, which uses a $m^*$ obtained from DFT calculations in the main text.

\subsection{DFT Calculations}

\begin{figure}[h!]
	\centering
	\includegraphics[width=0.75 \textwidth]{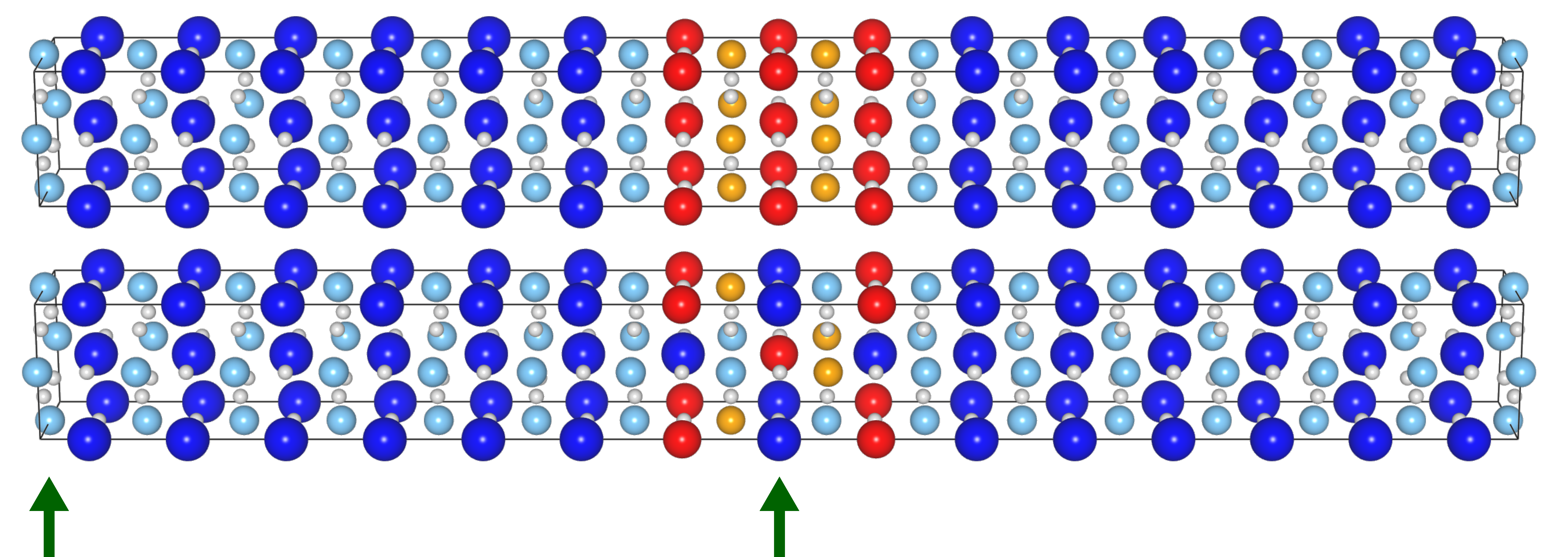}
	\caption{Unit cell of the superlattice structure: $\rm (STO)_{12}/(LAO)_{2}$ and $\rm (STO)_{12}/(LASTO$:$0.5)_{2}$. Two additional planes of $\rm TiO_2$ and $\rm LaO$ results in non-stoichiometry of the unit cell (green arrows).}\label{fig:superlattices}\end{figure}

{In this work, we study off-stoichiometric superlattices of (STO)$_\mathrm{m}$/(LAO)$_2$ and (STO)$_\mathrm{m}$/(LASTO:0.5)$_2$, for full-compensated interfaces. In contrast to previous DFT studies, we do not explicitly remove charges from the interface; instead, the charge density is directly tuned through the composition of LASTO:$x$. The geometry of the studied systems is illustrated in Fig.~\ref{fig:superlattices}, with $m = 12$ for clarity. We compute the structural and electronic properties of non-stoichiometric $\rm (STO)_{m}/(LAO)_{2}$ and $\rm (STO)_{m}/(LASTO$:$0.5)_{2}$ superlattices, with two $n$-type interfaces, and $m = 12,~22$ and $30$. The non-stoichiometry is set by adding an additional $\rm TiO_2$ and an additional $\rm LaO$ layers in their respective sub-lattice as shown in Fig.~\ref{fig:superlattices}. The superlattice model is known to mimic fully compensated interfaces, using the additional $\rm AlO_2$ planes to provide one single additional electron per 2D unit cell, which is shared by the two $n$-type interfaces, resulting in  $0.5~e^-/a^2$ manifested by the polar catastrophe scenario, i.e. $n_\mathrm{2D} = 3.3 \times 10^{14}~cm^{-2}$. Replacing the $\rm LAO$ layer with the LASTO:0.5 alloy only provide 0.5 additional electron in the cell, resulting in $0.25~e^-/a^2$ as predicted by the polar catastrophe scenario.}

{Periodic boundary conditions are used. To mimic the substrate epitaxial strain, the in-plane cell parameter is fixed to the cubic bulk value of STO, $a_\mathrm{STO}$ = 3.88~\AA, as determined in the framework of B1-WC. We use the tetragonal $P4/mmm$ space group for the superlattice with LAO interlayer, and $\displaystyle P\overline{4}m2$ space group for the superlattice with $\rm LASTO$:$0.5$ interlayer, with an in-plane $\displaystyle \sqrt{2}a \times \sqrt{2}a$ supercell approach to accomodate the solid solution. In this manner, each supercell contains $2\times (5[m + n + 1])$ atoms.}

{The present section aims to fully detail the results of our DFT calculations. As the  superlattices are symmetric, the following results are only displayed for only one half of each superlattice.}

\subsubsection{Structural properties}

\begin{figure}[ht!]
	\centering
	\includegraphics[width=0.75 \textwidth]{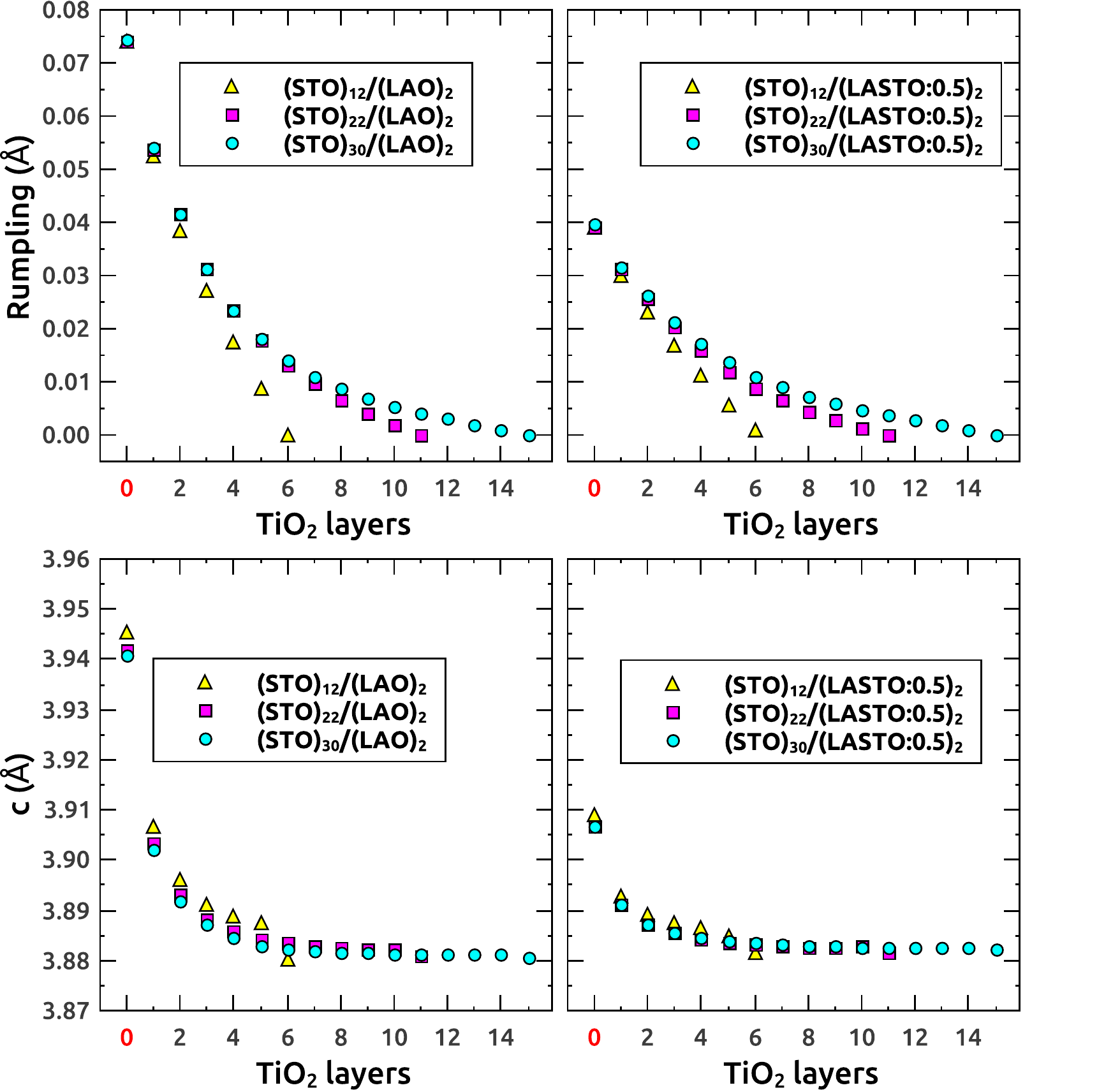}
\caption{Rumpling in $\rm TiO_2$ plane in STO (0 is the $\rm TiO_2$ interface), and out-of-plane lattice parameter $c$ of STO.}\label{fig:data}\end{figure}
{The structural properties of the $\rm STO$ substrate are summarized in Figure~\ref{fig:data}, where the rumpling in $\rm TiO_2$ planes (labeled from 0 to 15, with 0 being $\rm TiO_2$ plane at the interface) and lattice parameter $c$ are plotted, for the different superlattices. The rumpling is defined as the distance between cation and anion planes within a single atomic layer along [001] direction, while $c$ is calculated as the distance between cation-anion median planes.} 

\begin{figure}[h]
	\centering
	\includegraphics[width=0.75 \textwidth]{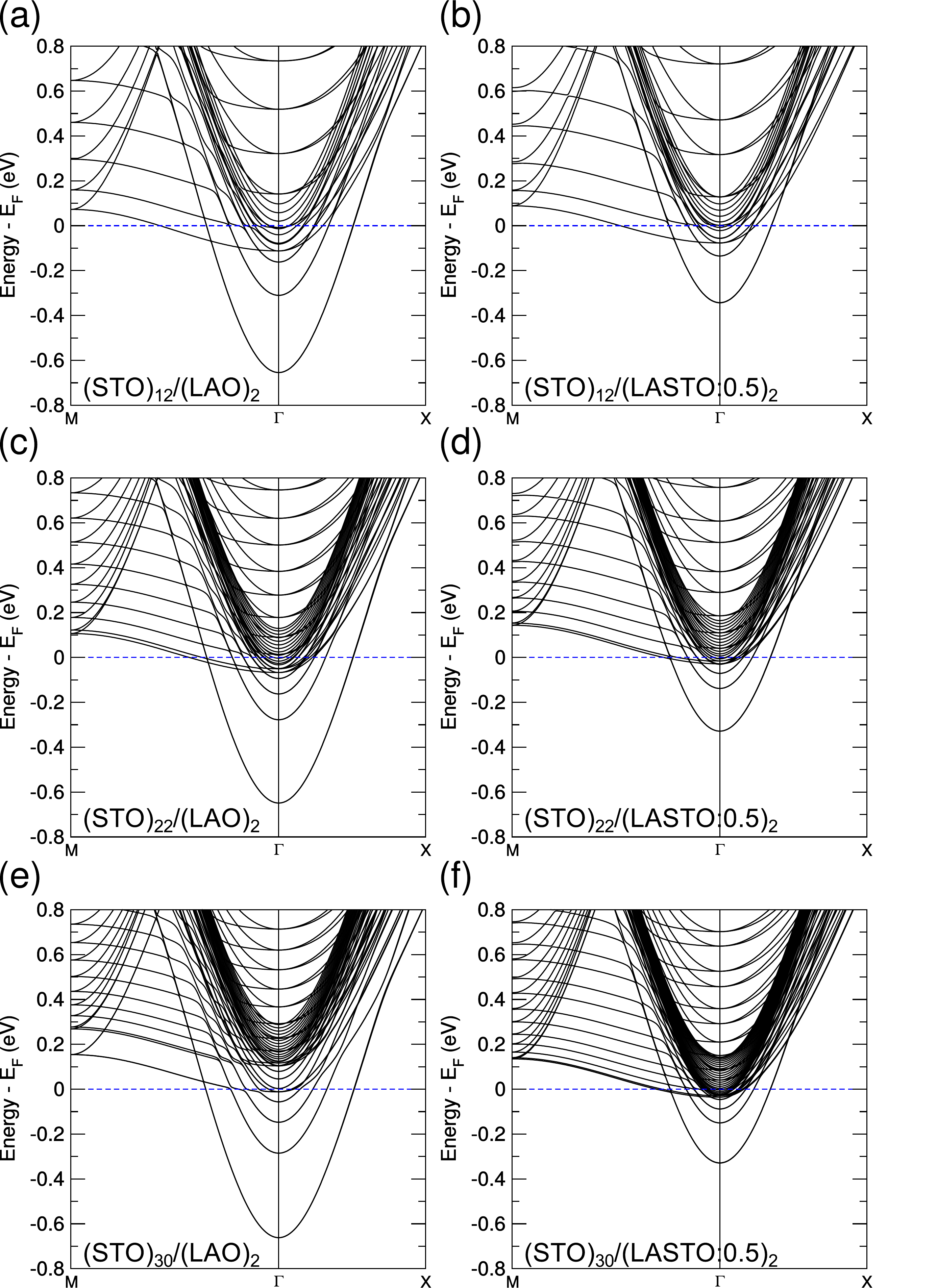}
\caption{Electronic band structure for (STO)$_m$/(LAO)$_2$ and (STO)$_m$/(LASTO:0.5)$_2$ superlattices with $m = 12$, $22$ and $30$. The Fermi level is shown by the horizontal dashed line.}\label{fig:band_structure}
\end{figure}

{For all superlattices, the first observation is that $c$ converges quickly to the lattice parameter of cubic bulk STO (determined by B1-WC). The second observation is that the rumpling also converges quickly toward 0~\AA, retrieving $m\overline{3}m$ point group symmetry of bulk STO. The finite size of STO imposes a value of the rumpling in the last $\rm TiO_2$ layer to be 0~\AA, independently of the superlattice size, and the last value of $c$ to be equal to $a$. The out-of-plane strain applied to the individual cells of STO and the value of the rumpling depend on the amount of transferred charges.}

\subsubsection{Electronic properties}

\begin{figure}[h]
	\centering
	\includegraphics[width=0.75 \textwidth]{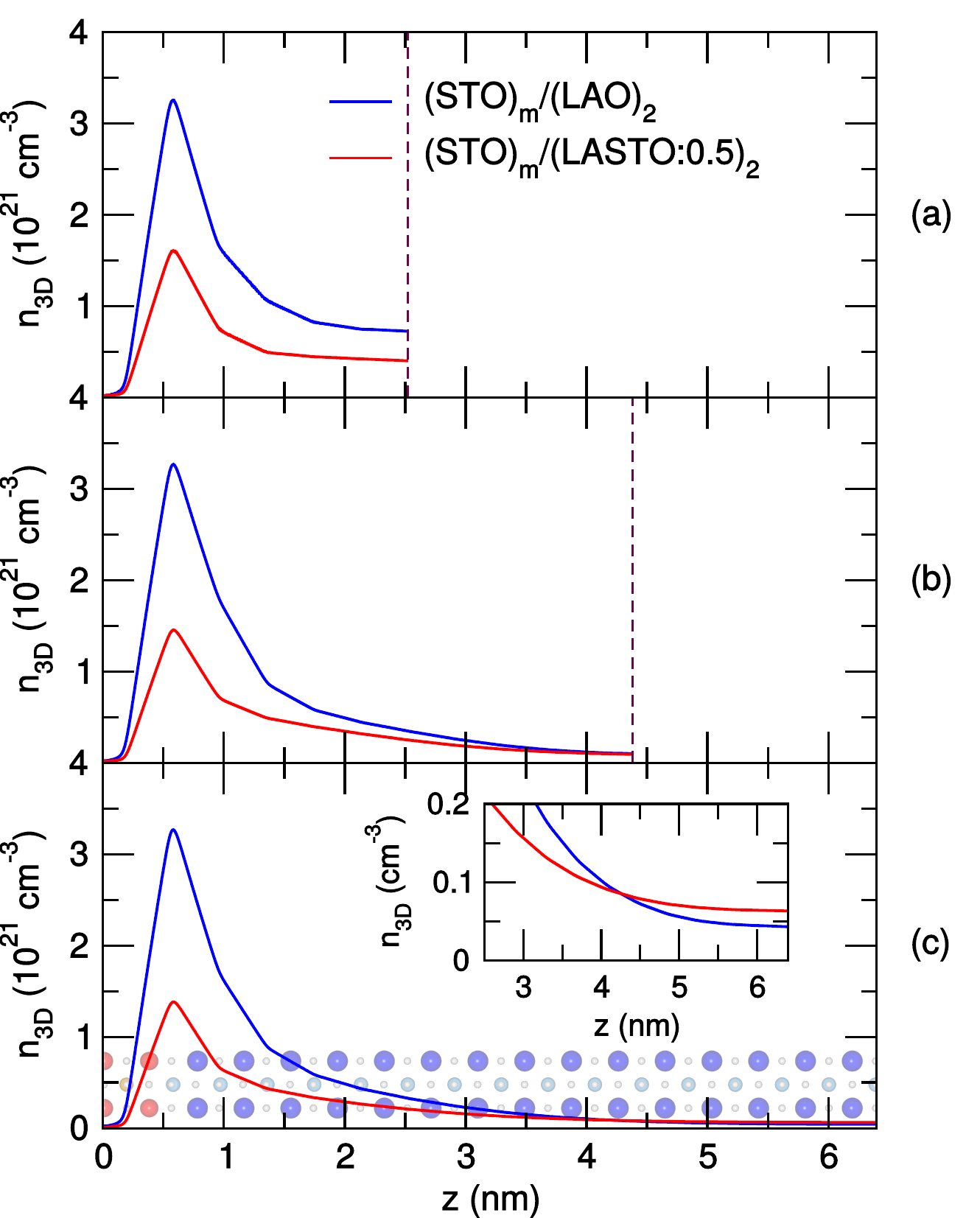}
\caption{Average macroscopic 3D electron density in (STO)$_m$/(LAO)$_2$ ((STO)$_m$/(LASTO:0.5)$_2$) superlattices for (a) $m = 12$, (b) $m = 22$ and (c) $m = 30$. The origin along $z$ is taken as the middle of each superlattice. The vertical dashed lines indicate the middle of the STO sublattice.}\label{fig:3DdensityDFT}
\end{figure}

\begin{figure}[h]
	\centering
	\includegraphics[width=0.75 \textwidth]{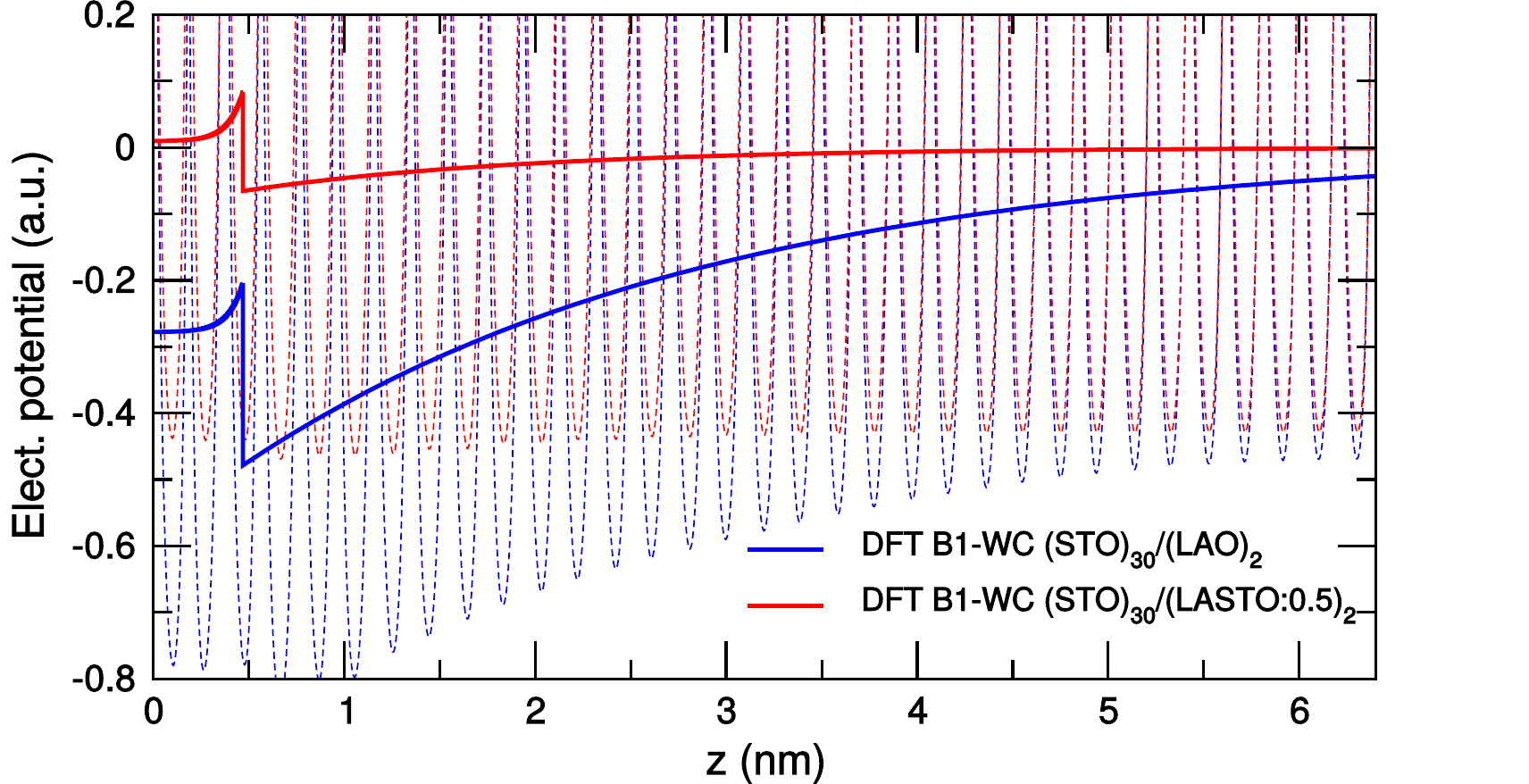}
\caption{Microscopic and averaged electrostatic potential in $\rm (STO)_{30}/(LAO)_2$ and $\rm (STO)_{30}/(LASTO$:$0.5)_2$.}\label{fig:potentialDFT}
\end{figure}

{To get insights on the electronic properties, the electronic band structures for all superlattices are calculated, as illustrated in Fig.~\ref{fig:band_structure}. The conduction band bottom is composed of ${\rm Ti}- d_{xz}$ and ${\rm Ti}-d_{xz}/d_{yz}$ states. The profile of the first available conduction states below Fermi level $E_\mathrm{F}$, which is the first populated ${\rm Ti} - d_{xy}$ close to the interface, does not significantly evolve with $m$. Moving from $\rm LAO$ to $\rm LASTO$:$0.5$ superlattices, the distance between the conduction band minimum and $E_\mathrm{F}$ changes from $\rm 0.65~eV$ to $\rm 0.33~eV$, which results in a change in electron density at the interface. For electrons occupying states far from the interface, only those that rest very close to $E_F$ are relevant. Here $m$ is found to be an important parameter: the thicker the $\rm STO$ slab in the superlattice is used, the denser the subband structure shows when close to $E_F$.}

\begin{table}[ht!]
\vspace{1em}
\centering
\caption{Mulliken decomposition of the transferred charges in the STO ${\rm Ti}-3d$ orbitals for superlattices with $m = 30$. In this picture, the remaining transferred charges, which are not found in ${\rm Ti}-3d$ orbitals, are located in the ${\rm O}-2p$ states within $\rm TiO_2$ planes.}\label{tab:mullikendecomposition}
\vspace{10px}
\begin{tabular}{ccc m{1em} cc}
\hline
\hline
& \multicolumn{2}{c}{\textbf{LAO}} && \multicolumn{2}{c}{\textbf{LASTO:0.5}}\\
\textbf{Ti} & $\textbf{d}_{\textbf{xy}}$ & $\textbf{d}_{\textbf{xz}}+\textbf{d}_{\textbf{yz}}$ && $\textbf{d}_{\textbf{xy}}$ & $\textbf{d}_{\textbf{xz}}+\textbf{d}_{\textbf{yz}}$\\
\hline
15        & 0.001 & 0.002 && 0.001 & 0.002\\
14        & 0.001 & 0.002 && 0.001 & 0.002\\
13        & 0.001 & 0.002 && 0.001 & 0.002\\
12        & 0.001 & 0.002 && 0.001 & 0.002\\
11        & 0.001 & 0.002 && 0.001 & 0.002\\
10        & 0.001 & 0.002 && 0.001 & 0.002\\
9         & 0.001 & 0.004 && 0.001 & 0.004\\
8         & 0.002 & 0.004 && 0.001 & 0.004\\
7         & 0.002 & 0.006 && 0.002 & 0.004\\
6         & 0.004 & 0.008 && 0.002 & 0.006\\
5         & 0.006 & 0.010 && 0.004 & 0.006\\
4         & 0.010 & 0.012 && 0.006 & 0.008\\
3         & 0.018 & 0.012 && 0.012 & 0.007\\
2         & 0.038 & 0.010 && 0.017 & 0.006\\
1         & 0.085 & 0.006 && 0.031 & 0.004\\
0         & 0.177 & 0.002 && 0.073 & 0.002\\
$\Sigma$  & 0.349 & 0.086 && 0.154 & 0.063\\
\hline
$\Sigma$ ${\rm Ti}-3d$ & \multicolumn{2}{c}{0.435} && \multicolumn{2}{c}{0.217} \\
$\%$     ${\rm Ti}-3d$ & 71$\%$ & 17$\%$ && 63$\%$ & 25$\%$\\
$\Sigma$ Rest & \multicolumn{2}{c}{0.060} && \multicolumn{2}{c}{0.032} \\
$\%$ Rest & \multicolumn{2}{c}{12$\%$} && \multicolumn{2}{c}{12$\%$} \\
\hline
\hline
\end{tabular}
\end{table}

{From the band structure, we calculate the effective masses, $m^*$, of ${\rm Ti}-3d$ bands. In bulk STO, the DFT B1-WC effective masses for the light and heavy bands are $m_l = 0.4~m_e$ and $m_h = 6.1~m_e$, respectively. In the superlattices, the values calculated for the light and heavy carriers at the interface are $m_l = 0.4~m_e$ and $m_h = 5.9~m_e$, which are very close to their bulk values.}
 
{The average 3D electron density of the total transferred charges $n_\mathrm{3D} (z)$ is plotted in Fig.~\ref{fig:3DdensityDFT} for (STO)$_{m}$/(LAO)$_{2}$ and (STO)$_{m}$/(LASTO:0.5)$_{2}$ ((a) $m = 12$, (b) $m = 22$ and (c) $m = 30$). $n_\mathrm{3D} (z)$ is calculated by averaging the electron density in the $ab$ plane, as moving along $z$. Several observations can be made from these figures:
(i) firstly, the density at the $\rm TiO_2$ interface does not evolve with increasing $m$, suggesting that the superlattice model properly captures the properties at the interfaces, as calculations performed in previous work. This is in consistency with the profile of the first occupied conduction band in STO, which does not change with $m$;
(ii) secondly, far from the interfacial $\rm TiO_2$ plane, $n_\mathrm{3D}$ converges very slowly with $m$. This is identified as size effect in the $\rm STO$ sublattice: the transferred electron charge cannot relax further beyond the boundary that is set by the size of the supercell (the boundary is indicated as the vertical dashed line);
iii) finally, substituting LAO with LASTO:0.5 decreases the $n_\mathrm{3D}$ at the interface, but electrons extend further, as illustrated in Fig.~\ref{fig:3DdensityDFT}c. The crossing of the two curves occurs at $\rm 4.25~nm$ from the middle of the superlattice.} 

{Despite the different sizes of STO in the superlattices, the $n_\mathrm{3D}$ never properly reaches 0. For $m = 30$, the minimum density (in $\rm (STO)_{30}/(LAO)_{2}$) in the $\rm STO$ sublattice is $\rm 4.3\times 10^{19}\rm~cm^{-3}$. For $m = 12$ or $22$, the de-confinement of the electrons far from the interface cannot be resolved. Hence, the results for $m = 30$ suggest that, if a STO with larger size is implemented in the calculation, we should observe a stronger de-confinement of the electrons from the interface for LASTO:0.5/STO system.}

{The confining potential in STO depends on the density of the total transferred charges. This is illustrated in  Fig.~\ref{fig:potentialDFT}, in which the electrostatic potential is plotted for both $\rm (STO)_{30}/(LAO)_{2}$ and $\rm (STO)_{30}/(LASTO$:$0.5)_{2}$ interfaces. The electric field in STO decreases moving from charge density of $0.5$ to $0.25~e/a^2$. The confinement of the transferred electrons is therefore also reduced.} 

{To have a complete grasp on the occupation of the different ${\rm Ti} - 3d$ orbitals, we list in Table~\ref{tab:mullikendecomposition} the orbital-resolved Mulliken population of the transferred electrons for both $\rm (STO)_{30}/(LAO)_{2}$ and $\rm (STO)_{30}/(LASTO$:$0.5)_{2}$ interfaces. Within this scheme, $88\%$ of the total electrons occupy ${\rm Ti} - 3d$ orbitals. The rest is attributed to ${\rm O} - 2p$ states. The different populations reflect the occupation of the associated band in Fig.~\ref{fig:band_structure}. The $d_{xy}$ population decreases very quickly from the first two interfacial $\rm TiO_2$ planes into STO. The tail of charge distribution is mainly dominated by electrons with $d_{xz}+d_{yz}$ character. When the density of the total transferred electrons is reduced, the relative proportion of the $d_{xz}+d_{yz}$ slightly increases.}

\bibliographystyle{Classes/naturemag}
\bibliography{References.bib}

%